\documentclass[11pt]{article}
\usepackage{graphicx} 
\usepackage{float}
\usepackage{rotating, graphicx}
\usepackage{booktabs}
\usepackage{multirow}
\usepackage{longtable}
\usepackage{array}
\usepackage{makecell} 
\usepackage{placeins}
\usepackage{caption}
\usepackage{subcaption}
\usepackage[a4paper,left=2cm,right=2cm,top=2.5cm,bottom=3cm]{geometry}
\usepackage[natbibapa]{apacite}
\usepackage{multibib}
\usepackage{xcolor}
\usepackage{url}
\usepackage{etoolbox}
\usepackage{appendix}
\usepackage{amsmath}
\usepackage{threeparttable}
\usepackage{etoc}
\usepackage{pdflscape}

\usepackage{tikz}

\newif\ifinappendix
\inappendixfalse

\AtBeginEnvironment{appendices}{\inappendixtrue}
\AtEndEnvironment{appendices}{\inappendixfalse}

\makeatletter
\let\origaddcontentsline\addcontentsline
\renewcommand{\addcontentsline}[3]{%
  \origaddcontentsline{#1}{#2}{#3}%
  \ifinappendix
    \ifstrequal{#1}{toc}{%
      \origaddcontentsline{atoc}{#2}{#3}%
    }{}%
  \fi
}

\newcommand{\tableofappendices}{%
  \begingroup
    \setcounter{tocdepth}{1}
    \@starttoc{atoc}%
  \endgroup
}
\makeatother

\begin{document}
\begin{titlepage}
    \LARGE
    \begin{center}
        \textbf{Tracks to Modernity: Railroads, Growth, and Civic Institutions in Denmark$^*$} \\
        
        \vspace{0.5cm}
        \large
        Tom Görges, TU Dortmund University \\ 
        Magnus Ørberg Rove, Statistics Denmark\\
        Paul Sharp, University of Southern Denmark \\ 
        Christian Vedel,$^1$ University of Southern Denmark \\ 
        \small
        
        \vspace{1.25cm}
        \large
        \textbf{Abstract} \\ 
        \vspace{0.75cm}
        \normalsize

    \parbox{0.80\textwidth}{
            We examine how railroads shaped Denmark’s nineteenth-century economic transformation and the diffusion of Grundtvigian institutions. Using a panel of 1589 parishes and a difference-in-differences design for staggered adoption, we find that railroad connection increased local population by 7 percent, partly through internal in-migration, and structural change away from agriculture. Railroad access also brought cultural and institutional change by increasing the local densities of folk high schools and community houses, though these results are less robust than the economic findings.
            
        }

        \vspace{0.025cm} 
        \parbox{0.80\textwidth}{
            \begin{flushleft}
                \textbf{Keywords:} Railroads, Economic Development, Institutional Change
            \end{flushleft}  
        }
    \end{center}
    \vfill
    
    \footnotesize
    $^*$We would like to thank Christian Nielsen for conscientious research assistance, and Nina Boberg-Fazlic for valuable feedback. This project was presented at the 2023 HEDG Mini Workshop (University of Southern Denmark), the 5th Annual Meeting of the Scandinavian Society for Economic and Social History (BI Norwegian Business School, Oslo), Microdata in Economic History (CES Munich), the Rome Workshop on Transport and History (Tor Vergata, Rome), and the World Economic History Congress 2025 (Lund). We would like to thank participants for insightful questions and comments. We are also deeply grateful to the editor, Dan Bogart, and the two referees, whose careful and constructive engagement with the manuscript significantly improved the paper. The codebase behind this paper as well as the paper itself benefited from improvements suggested by ChatGPT and Claude.\\
    All code required to reproduce the analysis from raw data through the final regressions is available at \\ \url{https://github.com/christianvedels/Tracks_to_modernity} \\
    $^1$ Corresponding author: Christian Vedel, christian-vs@sam.sdu.dk
\end{titlepage}

\section{Introduction}
In the late nineteenth century, Denmark underwent a profound transformation. Once a largely agrarian economy, the country rapidly industrialized and developed inclusive institutions that laid the foundation for its modern prosperity. A central driver of this transformation was the cooperative movement, particularly in dairy production, which revolutionized agricultural productivity and created a model of collective economic organization that persisted into the twentieth century \citep{milkandbutter,BobergFazlic2023}. But what made this development possible? While existing research has highlighted the importance of factors such as access to capital, technological innovation, and institutional frameworks, a crucial piece of the puzzle remains underexplored: the role of railroads. Did railroad expansion merely facilitate economic growth, or did it also serve as a catalyst for deeper institutional shifts?

The present work examines whether railroads not only promoted economic expansion but also played a fundamental role in shaping Denmark's institutional landscape. Specifically, we ask: \textit{Did the expansion of Denmark's railroad network in the nineteenth century contribute to both economic development and the spread of civic institutions, particularly community houses and folk high schools?} These institutions, closely linked to the Grundtvigian movement, a revivalist Lutheran movement central to Danish national identity \citep{bentzen2023holy,bentzen2024assimilate}, played a critical role in fostering civic engagement, education, and collective decision-making — key prerequisites for the rise of cooperative enterprises. By analyzing the link between railroad access, development, and the emergence of these institutions, we provide new insights into how infrastructure development shaped both markets and ideas.

To answer this question, we employ a historical panel dataset that links railroad expansion to economic and institutional outcomes at the parish level. We use a difference-in-differences framework to estimate the causal effect of railroad access on population growth, occupational structure, and the spread of community houses and folk high schools. Recognizing concerns about heterogeneous treatment effects and staggered adoption, we apply the method of \citet{Callaway2021} to ensure that our estimates accurately capture the impact of rail connectivity over time.

Our results show that railroad expansion had a significant and persistent impact on economic and social outcomes. First, we find that parishes connected to the railroad experienced substantial population growth, confirming findings from other contexts, such as Sweden \citep{berger_enflo_2017} and Prussia \citep{hornung_2015}, where railroads drove long-term demographic shifts. We further show that this population growth was partly driven by increased internal migration, as railroad access attracted migrants from other parts of the country. This growth was not merely a matter of redistribution; railroad access also accelerated structural change and increased employment in manufacturing and other non-agricultural sectors, consistent with evidence from England and Wales \citep{bogart2022} and the United States \citep{donaldson2016, atack2011railroads, hornbeck24manu}.

Second, we document a positive relationship between railroad access and the spread of Grundtvigian institutions. Community houses and folk high schools---key venues for political discourse and collective organization---were significantly more likely to emerge in areas exposed to the railroad. Folk high schools clustered at railroad nodes and their immediate surroundings, consistent with their need for a wider catchment area and inflows of students from distant parishes. By contrast, community houses were no more likely to locate in railroad nodes than in their rural hinterlands, reflecting their function as locally oriented institutions: they did not require rail access to operate, but rail connectivity made it more likely that Grundtvigian ideas arrived and that local establishment followed. This aligns with research showing that railroads facilitated the diffusion of social movements, such as the temperance movement in the U.S. \citep{García-Jimeno_et_al_22} and grassroots political mobilization in Sweden \citep{Melander_2020}. Our findings suggest that, beyond economic integration, railroads played a role in diffusing Grundtvigian institutions and the civic participation they enabled.

Our study contributes to several strands of literature. First, it builds on the extensive research on railroads and economic development by showing that their impact extended beyond market access and trade expansion to shaping institutional change. While prior work has emphasized the role of railroads in industrialization \citep{atack2011railroads, Atack2022}, we highlight their role in fostering the social infrastructure necessary for Denmark's cooperative movement. Second, we contribute to the literature on institutional change and political economy by demonstrating that physical infrastructure can serve as a conduit for the diffusion of civic institutions and organizational forms. Finally, our findings contribute to the broader literature on economic history by providing empirical evidence on the long-debated relationship between market access, institutional development, and economic modernization.

The paper proceeds as follows. Section 2 reviews the related literature, situating our study within the broader research on railroads, economic development, and institutional change. Section 3 outlines the conceptual framework, linking railroad access to economic change and the diffusion of Grundtvigian institutions, and motivating the parallel analysis of economic and institutional outcomes. Section 4 provides historical background on Denmark's railroad expansion and discusses the historical mechanisms linking rail connectivity to economic and civic-institutional change. Section 5 describes the data sources and empirical strategy, detailing our approach to estimating the causal effects of railroad expansion. Section 6 presents the main results, examining the impact of railroads on population growth, structural change, and the diffusion of community houses and folk high schools. Finally, Section 7 concludes with broader implications for understanding how infrastructure investments contribute to both economic modernization and social transformation.

\section{Literature Survey}
The role of railroads in economic development has long been a matter of debate. \citet{fogel1964railroads} argued that railroads were not essential to U.S. economic growth, as waterways and canals could have served as viable substitutes. This view challenged earlier narratives that positioned railroads as the driving force behind nineteenth-century American development. However, more recent work has provided strong evidence of their transformative impact. For instance, \citet{atack2010} document that while rail access had only a modest effect on population density, it was instrumental in driving urbanization in the American Midwest between 1850 and 1860. Moreover, \citet{donaldson2016} show that railroads significantly expanded market access for U.S. producers, leading to substantial increases in agricultural land values. Finally, \citet{hornbeck24manu} demonstrate that railroads contributed far more to aggregate productivity growth than previously estimated, particularly when accounting for inefficiencies in market access. Taken together, these findings underscore that railroads were not merely an alternative transportation mode but a crucial factor in shaping market integration and long-term economic growth, a perspective that guides our analysis of nineteenth-century Denmark below.

Beyond the United States, extensive research has examined the relationship between railroads and population growth. Several studies highlight that early access to railroads led to lasting changes in settlement patterns and economic activity. \citet{berger_enflo_2017} show that in Sweden, towns that gained early railroad access experienced persistent population growth. However, towns that were connected later did not necessarily catch up, suggesting limited convergence effects. Similarly, \citet{hornung_2015} finds that railroad access significantly boosted urban growth in nineteenth-century Prussia, underscoring the role of transport infrastructure in shaping urbanization. The long-term effects of railroad infrastructure are also evident in cases where rail access was later removed. \citet{gibbons2024} show that the large-scale closure of railroad lines in Britain had persistent negative consequences. Studies from other European contexts provide further evidence of railroad-driven population growth.\footnote{\citet{Esteban-Oliver_2023} for Spain, \citet{braun2022} for Württemberg, \citet{koopmans2012} for the Netherlands} However, beyond the Danish historical literature including that cited below, no comparable evidence exists for Denmark on its own, with the exception of \citet{vedel2024perfectstorm} on the related case of waterways, leaving open the question of how railroads shaped Danish settlement patterns.\footnote{\citet{enflo2018transportation} and \citet{AlvarezPalau2021} do provide Pan-Nordic and Pan-European evidence.}

While railroads influenced population dynamics, their impact extended beyond this. \citet{bogart2022} find that railroads also caused an occupational shift away from agriculture and \citet{berger2019} documents that railroad expansion in nineteenth-century Sweden played a crucial role in the transition from agricultural to industrial employment. Similarly, \citet{atack2011railroads} and \citet{hornbeck24manu} show that railroad access is associated with a shift into new forms of production. \citet{Korn_Lacroix_24} highlight within-sector reallocations, which they measure using bankruptcy data. \citet{Chiopris_2024} shows that the German railroad network increased specialization in knowledge production, raising the creation of new ideas while reducing their diffusion across fields. Together, this body of work shows that railroads reshaped local economies not only by expanding markets but also by altering the structure of employment itself. Our Danish results fit this pattern: rail access is associated with a rise in manufacturing and other non-agricultural occupations.

Another outcome emphasized in the literature is migration, with railroads serving as a facilitator of internal mobility. \citet{enflo2018transportation} highlight that railroads played a critical role in Nordic industrialization by enabling resource exports, supporting rural industry, driving migration toward expanding economic centers, and reducing regional inequality. \citet{Büchel_and_Kyburz_2018} show that population growth appears to have come at the cost of localized displacement. \citet{Mojica_Henneberg_2011} document how railroad expansion in France, Portugal, and Spain facilitated urban growth by attracting populations to cities while simultaneously accelerating rural depopulation. \citet{escamilla2024all} finds that railroad expansion in Mexico significantly facilitated mass migration to the US. Taken together, these studies underline that population effects often reflect spatial reallocation rather than pure aggregate growth---an interpretation that aligns closely with the Danish case we analyze below.

Railroads also had other important demographic effects, particularly on fertility. Expanding railroad networks slowed fertility decline by increasing market access, which raised incomes and thereby affected the quantity-quality trade-off \citep{Ciccarelli_et_al_WP}. In industrializing England and Wales, \citet{galofre2024railroads} finds that railroad expansion increased local fertility rates by approximately three percent, suggesting that improved economic conditions and connectivity may have encouraged larger families. In contrast to these findings, we find no evidence that railroad expansion influenced fertility in Denmark.

Railroads also played a crucial role in enabling social movements. \citet{Melander_2020} finds that the expansion of Sweden's railroad network between 1881 and 1910 facilitated the spread of grassroots social movements by increasing individual mobility and connectivity, leading to faster membership growth, more organizations, and greater political mobilization in the 1911 election. Similarly, \citet{García-Jimeno_et_al_22} show that railroads and the telegraph helped spread the 1873–1874 Temperance Crusade in the US, highlighting how traditional communication networks enhanced the reach and organizational capacity of social movements. Beyond the European and North American contexts, \citet{brooke2018social} demonstrate that in interwar Egypt, the railroad network also facilitated the diffusion of the Muslim Brotherhood, one of the first organized Islamist movements. Their findings underscore that improved transport and communication infrastructure can enable the spread of a wide range of ideological currents, not only liberal or democratic ones. Taken together, these studies form a broader literature linking improved communication infrastructure to political mobilization. Our findings extend this literature by showing that in the Danish case, railroads facilitated the diffusion of Grundtvigian institutions---community houses and folk high schools---thereby contributing to a broader cultural shift toward more inclusive local institutions.

\citet{Andersson_et_al_2023} show that the expansion of the Swedish railroad network facilitated a market for ideas by reducing communication and transport costs, enabling inventors to develop and commercialize innovations beyond their local economies. In Italy, however, the link between railroads and innovation was delayed. \citet{MartinezNuvolariVasta2024} find that railroad expansion spurred innovation only after several decades and primarily benefited independent inventors and low-quality patents, with the most pronounced effects occurring during the state-driven first wave of railroad construction (1861–1878).  Other studies emphasize broader economic, social, and educational impacts of the railroad. In Sweden, \citet{Cermeño_Enflo_Lindvall_2022} show that nineteenth-century railroads helped school inspectors monitor education, leading to higher attendance rates and an increased emphasis on nation-building subjects, while more remote schools remained under local religious control. In England and Wales, \citet{costas2020train} find that railroad access increased intergenerational mobility. Additionally, railroad expansion has been linked to improvements in literacy in India \citep{Chaudhary_Fenske_2023} and rising school enrollment in the U.S. \citep{atack2012impact}, underscoring how transport infrastructure influenced human capital accumulation.  

Overall, the existing literature highlights the transformative role of railroads in shaping economic and social development across diverse historical contexts. Building on this, we examine the role of railroads in Denmark's nineteenth-century transformation, with a particular focus on both economic and institutional change. While previous research has extensively documented how railroads influenced population growth, industrialization, and market integration, less attention has been paid to their role in shaping social institutions and cultural movements. No existing study links railroads to the rise of Grundtvigian institutions, despite their centrality to Denmark's distinctive path toward inclusive local organization. By analyzing the relationship between railroad expansion and the spread of Grundtvigian institutions, we contribute to a growing body of work that considers the broader societal impacts of transport infrastructure and highlight a mechanism---railroads as conduits for civic and cultural diffusion.

\section{Conceptual Framework}

We argue that rail connection sets in motion two complementary processes: (i) improved market access and mobility, which affects local economic development; and (ii) faster diffusion of ideas and organizational forms, which affects the spread of Grundtvigian institutions. We therefore study economic and institutional outcomes in parallel.

\textit{Getting to Denmark} has become a widely adopted analogy describing the challenge of establishing modern, stable, and democratic institutions. \citet{fukuyama2011origins} emphasizes that such institutions do not emerge fully formed but rather develop through complex historical processes, requiring a stable balance of the state, rule of law, and accountable government. The case of Denmark exemplifies this gradual evolution. In the eighteenth century, Denmark was an absolutist and militarized state comprising multiple cultural and linguistic groups. The transition to a more cohesive and democratic society was far from predetermined. A central part of this shift was the emergence of Grundtvigian ideas, associated with N.F.S. Grundtvig, whose emphasis on education, civic engagement, and national identity helped reshape Denmark's institutional and cultural landscape \citep{BobergFazlic2023,Fukuyama2015}.

Grundtvigianism, emerging in the mid-nineteenth century, played a crucial role in fostering democratic participation and rural enlightenment. The movement promoted education through folk high schools, which emphasized lifelong learning, civic responsibility, and national identity. These institutions provided an ideological foundation for the cooperative movement, which became a hallmark of Denmark's economic model. Similarly, \citet[p. 68]{Korsgaard_2008} notes that the folk high schools operated ``as an institution of formation as well as an educational body - not only [for] skill training, but also [for] creating national citizens''. He further argues that this role would have ``produced [the] mental and cultural conditions that greatly contributed to the Danish peasantry's choice of the co-operative model''.\footnote{Korsgaard here draws on his earlier analysis in \citet[pp. 191-208]{Korsgaard_1997}.} Alongside these schools, community houses (\textit{forsamlingshuse}) were established as venues for public discourse, education, and political organization. Together, these institutions reinforced a participatory ethos that underpinned Denmark's political and economic modernization \citep{BobergFazlic2023,bentzen2023holy,bentzen2024assimilate}.

While both folk high schools and community houses reflect Grundtvigian influence, they capture different mechanisms. Folk high schools were typically founded through organized, often elite-led initiatives and diffused ideas through structured education---a more top-down channel. Community houses emerged from grassroots efforts and anchored everyday participation and decision-making---a more bottom-up channel. We therefore treat them separately in the empirical analysis, with folk high schools reflecting a more organized channel of Grundtvigian diffusion, and community houses reflecting the local uptake of these ideas through participatory organization.

Shortly after the emergence of these institutions, cooperative creameries came to play a crucial role in Denmark's transformation \citep{BobergFazlic2023}. These cooperatives not only facilitated agricultural modernization but also reinforced inclusive institutions that underpinned long-term development. However, their formation required both tangible and intangible preconditions. On the practical side, successful creameries depended on key inputs such as centrifuges, a reliable supply of high-quality milk, and access to energy sources like coal. Just as importantly, cooperatives required social and institutional foundations---a space for deliberation and decision-making (the constitutional general assembly) and the conceptual framework for collective organization \citep{milkandbutter}.

Figure \ref{fig:conceptual_framework} summarizes our conceptual framework. Railroads shape economic outcomes and facilitate the diffusion of Grundtvigian institutions. The two-headed arrow between economic change and institutional change reflects the possibility that each reinforced the other---rising incomes could sustain the organizational capacity needed to build institutions, while institutions could in turn promote economic participation. We do not attempt to decompose these indirect pathways; our empirical strategy estimates the reduced-form effect of railroad connection on each set of outcomes in parallel.

\begin{figure}[ht]
    \centering
    \caption{Conceptual framework: Railroads, Development, and Modern Denmark}
    \scalebox{0.95}{\usetikzlibrary{positioning}

\definecolor{own_blue}{HTML}{328CC1}
\definecolor{greyblue}{HTML}{5A7487}
\definecolor{lightgreyblue}{HTML}{8498A5}

\begin{tikzpicture}[
    node distance=1cm and 1.5cm,
    box/.style={
      rectangle,
      draw = black,
      line width = 1pt,
      rounded corners,
      text width=6.2cm,
      minimum height=1cm,
      align=center,   
      inner sep=6pt,  
    }
]

\node[
box, 
] (rail) {Railways};

\node[
box, below= 1cm of rail 
] (econ) {
    i. Economic Change\\
    \vspace{-0.5em}
    \begin{flushleft}
    \small
    \hspace{0.5em} -- Population Growth\\
    \hspace{0.5em} -- Fertility\\
    \hspace{0.5em} -- Industrial employment\\
    \hspace{0.5em} -- Non-agricultural employment\\
    \hspace{0.5em} -- Socio-economic status \\
    \hspace{0.5em} -- Migration \\
    \end{flushleft}
    
    };

\node[
box, below right= of rail, xshift=1cm 
] (inst) {
    ii. Institutional Change \\ (Grundtvigianism)\\
    \vspace{-0.5em}
    \begin{flushleft}
    \small
    \hspace{0.5em} -- Folk High Schools\\
    \hspace{0.5em} -- Community Houses
    \end{flushleft}
      
    };

\node[
box, below=1cm of econ, 
] (modern) {Modern Denmark};

\draw[->, very thick] (rail) -- (econ);
\draw[->, very thick] (rail) -| (inst);
\draw[->, very thick] (econ) -- (modern);
\draw[->, very thick] (inst) |- (modern);
\draw[<->, thick, dashed] (econ.east) to[bend left=20] (inst.west);

\end{tikzpicture}

}\\
    \vspace{0.5em}
    \parbox{1\textwidth}{%
        \footnotesize \textit{Notes}: This figure summarizes the conceptual framework. Railroads affected both economic change and the diffusion of Grundtvigian institutions. The two-headed arrow between economic and institutional change indicates that these processes may have reinforced each other. We estimate the reduced-form effect of railroad connection on each set of outcomes; we do not attempt to decompose whether institutional effects ran through or independently of economic change.
    }
    \label{fig:conceptual_framework}
\end{figure}

\FloatBarrier
\section{Historical Background} \label{hist}
The Danish trunk railroad lines were constructed between 1847 and the mid-1870s, with the state playing a central yet evolving role in their development. As only the state could grant concessions for railroad construction, the design and expansion of the network were deeply intertwined with contemporary political debates. Consequently, the layout and timing of the railroads reflected political interests rather than purely economic considerations \citep{thestrup1997dampen, norgaardolesen1990}. Two competing perspectives shaped this debate. One faction, primarily consisting of Jutlandic peasants engaged in cattle exports, advocated for a north-south railroad linking Jutland with Hamburg and Husum to facilitate livestock trade. In contrast, the national-liberal movement sought to reduce Denmark's dependence on German markets, particularly Hamburg, by strengthening direct trade routes to England. Their preferred railroad layout emphasized east-west connections from Jutland to ports on the Limfjord and the western coast, ensuring efficient access to English markets while maintaining strong links to Copenhagen and key market towns \citep{hansen1972}.

Denmark's first railroad line, within the (then) personal union with Schleswig-Holstein, opened in 1844 between Kiel and Altona near Hamburg. Although proposals for a railroad linking the Baltic and North Seas dated back to 1831, the Danish government initially resisted such plans, as they threatened to bypass the lucrative Sound Toll, which accounted for over a quarter of state revenue in 1847. However, geopolitical pressures eventually forced Denmark's hand, leading to the construction of the Kiel-Altona railroad \citep{thestrup1997dampen}. At the same time, national-liberal proponents had suggested an alternative railroad from Flensburg to Tønning, which would have strengthened Denmark's connection with England. However, in an effort to maintain Holstein within the union, the Danish government ultimately favored the Kiel-Altona line. This pattern illustrates how railroad planning was subject to national priorities and political considerations rather than strictly economic efficiency \citep{lampe2015danes}.

In 1854, British entrepreneur Sir Morton Peto successfully proposed the Flensburg-Tønning line, adding a side track to Rendsburg to connect with the existing Kiel-Altona railroad. His vested interest in steamship operations between England and Tønning partly motivated this project \citep{thestrup1997dampen}. In modern-day Denmark, the first domestic railroad line opened in 1847, linking Copenhagen with Roskilde. By 1856, the extension to Korsør dramatically reduced travel times across the country, facilitating connectivity between Copenhagen and Altona in just 14 hours, compared with the 36-hour steamboat journey from Copenhagen to Kiel prior to rail construction \citep{buch1933danmarks}. Proposals for further expansion continued to reflect competing priorities. Plans emphasizing direct agricultural export routes received strong support from Jutlandic peasants, while national liberals criticized them for neglecting populous cities along Jutland's eastern coast and for failing to provide a direct connection to Copenhagen \citep{thestrup1997dampen}.

After years of political disagreement and economic uncertainty, a compromise was reached in 1861. The final plan reflected national-liberal priorities by including an east-west railroad from Aarhus to Limfjorden, enabling steamship connections to England via the newly emerged Agger Channel \citep{vedel2024perfectstorm}. The north-south railroad route was also adjusted to pass through the fortress town of Fredericia, fulfilling strategic military objectives following Denmark's experiences in the First Schleswig War. Additionally, a railroad across Funen, linking Nyborg and Middelfart, was approved, along with a direct line from Flensburg to Kolding, completing the north-south connection to Hamburg \citep{norgaardolesen1990}.

The Second Schleswig War in 1864, which resulted in Denmark's loss of southern Jutland, further reinforced the strategic shift towards direct trade with England. This transition was already underway in the 1850s, as evidenced by price integration between the Danish and British butter markets \citep{lampe2015danes}. The loss of Schleswig and Holstein accelerated this reorientation, making reliance on Hamburg politically and economically unviable. In response, Denmark prioritized the construction of Esbjerg harbor on Jutland's southwest coast, along with railroad connections from Holstebro, Varde, and Vejle. The harbor's rapid development reflected the new commercial reality, as it provided a crucial alternative to Hamburg for agricultural exports \citep{thestrup1997dampen}.

Denmark experienced significant economic growth from the late nineteenth century through the early twentieth century, shifting from a primarily grain-based economy to one focused on animal products. This transition was facilitated by the expansion of railroad infrastructure, which enabled efficient livestock transportation and supported the growing dairy industry. The railroad network, combined with easy access to the coast throughout the country, also allowed for the import of critical inputs in particular coal, which were essential for large-scale dairy production \citep{henriques2016danish}. Moreover, the spread of railroads contributed to the rise of new urban centers, often referred to as \textit{railroad towns} (\textit{stationsbyer}). These settlements emerged around railroad stations, developing as local transport and commercial hubs. Many of these towns hosted creameries, grain mills, and other industries tied to the agricultural sector, reinforcing Denmark's position as a leading exporter of dairy products. Langå, for example, became an important railroad junction, and its rapid development highlights how railroad connectivity spurred localized economic growth \citep{grothfertner2013}.

These railroad towns also provide a concrete setting for understanding how railroad access could shape civic institutions. By civic institutions, we mean the voluntary local organizations, meeting places, and schools that supported popular education, association life, and collective action. Contemporary accounts of Danish \textit{railroad towns} [stationsbyer] emphasize that stations commonly attracted a bundle of economic and social functions, including cooperative creameries, community houses, folk high schools, and agricultural schools \citep{stationsbyLex,historievejenStationsbyer}. Here, we use Grundtvigian institutions to refer to schools and associational spaces inspired by N.F.S. Grundtvig's emphasis on popular education and civic participation. These institutions were tied to the agricultural transformation described above. As agriculture shifted from grain to animal products, farmers needed places for processing, marketing, meetings, and instruction. The railroad did not mechanically produce the same institution everywhere; rather, it created nodal places where economic activity, mobility, and civic organization could reinforce each other.

Several local histories make this mechanism visible. St\o{}vring H\o{}jskole, a folk high school in Himmerland in northern Jutland, was founded in 1885 by Kristian \O{}sterg\aa{}rd and Ludvig Mosb\ae{}k. \O{}sterg\aa{}rd was a teacher and writer who became the school's first head; Mosb\ae{}k was a local gardener and seed trader whose nursery business supplied both the initial physical premises and the practical gardening component of the school \citep{stoevringBegyndte}. The school had first been considered for B\ae{}lum, a village in eastern Himmerland, but Mosb\ae{}k argued for St\o{}vring because it was ``by a railroad station in the middle of Himmerland'' and had available buildings that could be adapted for school use \citep{stoevringBegyndte}. The local archive describes how the Hobro--Aalborg line, opened in 1869, transformed St\o{}vring from a small rural village into a station town with shops, workshops, a cooperative creamery, a cooperative store, and a folk high school \citep{stoevringJernbane}. The case therefore shows that railroad access could factor directly into decisions about where to locate a school, while also anchoring the wider economic infrastructure around it.

Vr\aa{} H\o{}jskole, in Vendsyssel in northern Jutland, provides an even clearer example of relocation toward the railroad. The school began in 1872 in Stenum, a small village southwest of Vr\aa{}, on Kirstine Terkelsen's family farm. Kirstine and her husband J\o{}rgen Terkelsen, a young folk-high-school teacher, had both been pupils in the educational environment associated with Christen Kold, and together they founded the first Grundtvigian folk high school in Vendsyssel \citep{vrenstedVraa}. After the Jutland longitudinal line reached Vr\aa{} and Br\o{}nderslev, the founders decided to move closer to the railroad; the local history states that, as the school grew, the idea quickly arose ``to move the high school closer to the railroad'' \citep{vrenstedVraa}. In 1890 the school moved to the station town of Vr\aa{}, where it was placed west of the railroad line; the school's own history describes this as a move ``stone by stone'' from Stenum to Vr\aa{} \citep{vraaOm,hojskolehistorieVraa}. These cases illustrate why folk high schools might respond more strongly to railroad access than community houses: they were boarding schools that depended on a wider catchment area, visiting lecturers, and flows of students from outside the immediate parish. This interpretation is also consistent with official statistics from 1904/05 and 1905/06, which noted that ``growing prosperity and cheaper railroad transport'' had contributed to the increasing tendency of students to attend high schools farther from home \citep{yearbook1905}.

Community houses followed a different logic. They were locally funded and locally used, but they were embedded in the same station-town environment of new associations, meetings, and cooperative enterprises. Tommerup Stationsby, west of Odense on Funen, is a useful example. The main line across Funen opened in 1865, and the station was placed outside the old village on B\aa{}geg\aa{}rd's land. The station town then grew rapidly: local sources report an increase from 88 inhabitants in 1860 to 637 in 1911 \citep{tommerupStationsby}. In 1888 farmers from seven parishes established Tallerup Andelsmejeri, a cooperative creamery, at the corner of Stationsvej and Tallerupvej, reflecting the general shift from grain to livestock production \citep{tommerupStationsby,gausdal1998}. A year later, P. Preisler, a merchant from Odense, transferred a plot of land for Tommerup Stations Forsamlingshus, the town's community house. In 1911 Vilhelm Sp\o{}hr, a Tommerup merchant who was also active in the community house's management and audited its accounts, donated additional land that allowed the house to expand \citep{gausdal1998,tommerupStationsby}. In the words of a local history, ``the community house had, from the beginning, a popular-educational purpose.'' It hosted lectures by folk-high-school principals and was used by smallholders' associations, temperance groups, singing associations, teachers, youth organizations, political associations, craftsmen, the cooperative store, railroad employees, the lecture association, and the creamery itself \citep{tommerupStationsby,gausdal1998}. In this case the railroad did not simply place a community house at a station; rather, it helped create a new local center in which cooperative dairying and associational life developed together.

Other station towns show similar clustering. A new settlement emerged around F\aa{}revejle Station, in Odsherred in northwestern Zealand and on newly reclaimed land at the Lammefjord, and became a center for mission schools, a folk high school later converted to a nursing school, a Grundtvigian free and continuation school (\textit{friskole} and \textit{efterskole}), a cooperative creamery, and other social institutions \citep{farevejleStation}. Jyderup, in western Zealand on the Roskilde--Kalundborg line from 1874, similarly developed workers' housing, a mission house, schools, and social institutions; the nearby villa S\o{}lyst later served as a women's home and a children's home, and today serves as a folk high school \citep{jyderupStation}. Additional cases point in the same direction: T\o{}ll\o{}se H\o{}jskole, in northwestern Zealand, was built at the eastern end of a station town on the Nordvestbanen [Northwest Line]; Ry H\o{}jskole, near Silkeborg in eastern Jutland, was located opposite Ry station and incorporated an earlier community house; and at S\ae{}rslev, on northwestern Funen, a community house was built next to the high school when existing meeting rooms became too small \citep{toelloeseHojs,ryHojs,saerslevMindeskrift}. We do not use these examples as evidence of causal effects in themselves. Instead, they clarify the historical mechanisms that motivate the empirical analysis. Folk high schools represent a more organized channel of diffusion, where access to a wider railroad-connected catchment made siting at or near railroad nodes attractive. Community houses represent the local uptake of civic and educational ideas, often in places where the railroad had already intensified economic activity and associational life. The distinction maps directly onto the empirical patterns below: railroad access should matter most for the location of folk high schools, while the diffusion of community houses may be more visible in local density and spillovers than in parish-level siting.

Other examples illustrate how railroads supported economic development through improved access to inputs and markets. On Funen, railroads distributed imported coal to rural producers and expanded markets for brick production around Stenstrup \citep[pp.~74--76]{fransen1997transport}. In southeastern Zealand, a dedicated railroad connecting the Faxe limestone quarries to the coast, together with new harbor facilities, supported expanding production and the growth of Faxe Ladeplads \citep{mikkelsen2020faxe}. These developments depended on complementary conditions: some Danish rural towns began growing before receiving railroads, while others developed without rail connections \citep{thogersen2013stationstiden}.

\begin{figure}[ht]
    \caption{Danish Railroad Expansion}
    \centering
    \begin{minipage}[b]{0.4\textwidth}
        \centering
        \includegraphics[width=\linewidth, trim={5cm 1cm 4.6cm 0.75cm}, clip]{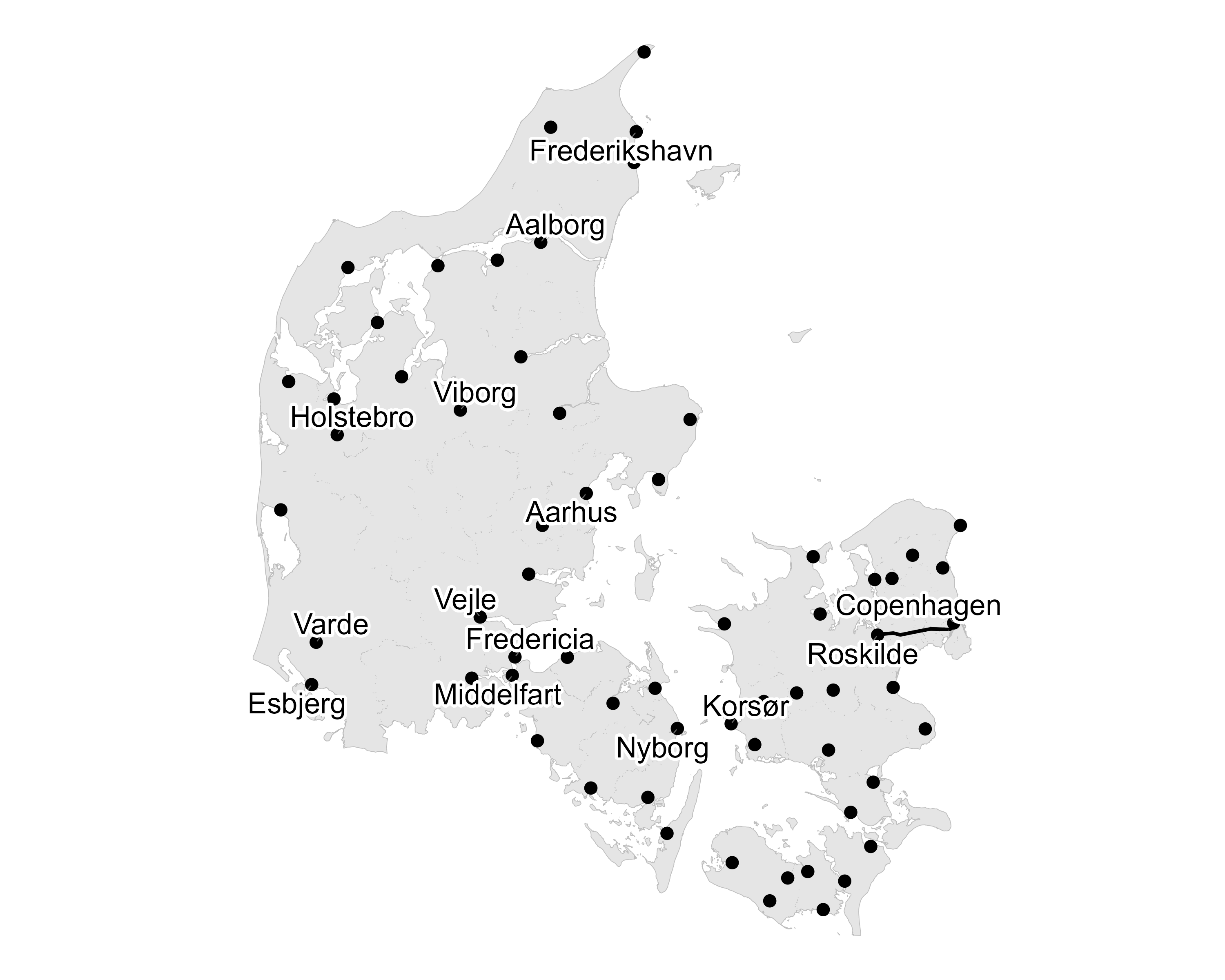} 
        \caption*{(A) 1850}
    \end{minipage}
    \hspace{2cm}
    \begin{minipage}[b]{0.4\textwidth}
        \centering
        \includegraphics[width=\linewidth, trim={5cm 1cm 4.6cm 0.75cm}, clip]{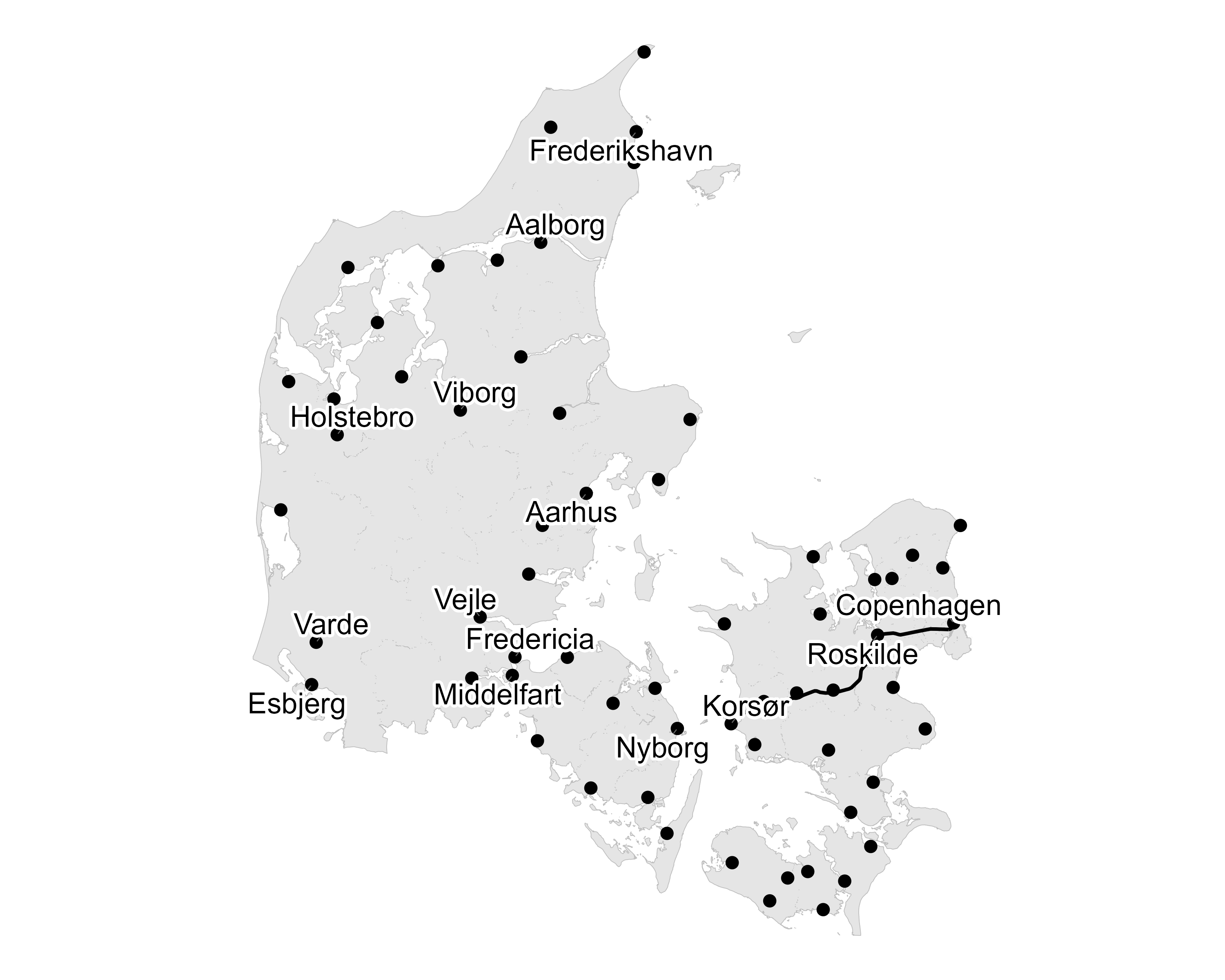} 
        \caption*{(B) 1860}
    \end{minipage}
    \vspace{1cm}
    \begin{minipage}[b]{0.4\textwidth}
        \centering
        \includegraphics[width=\linewidth, trim={5cm 1cm 4.6cm 0.75cm}, clip]{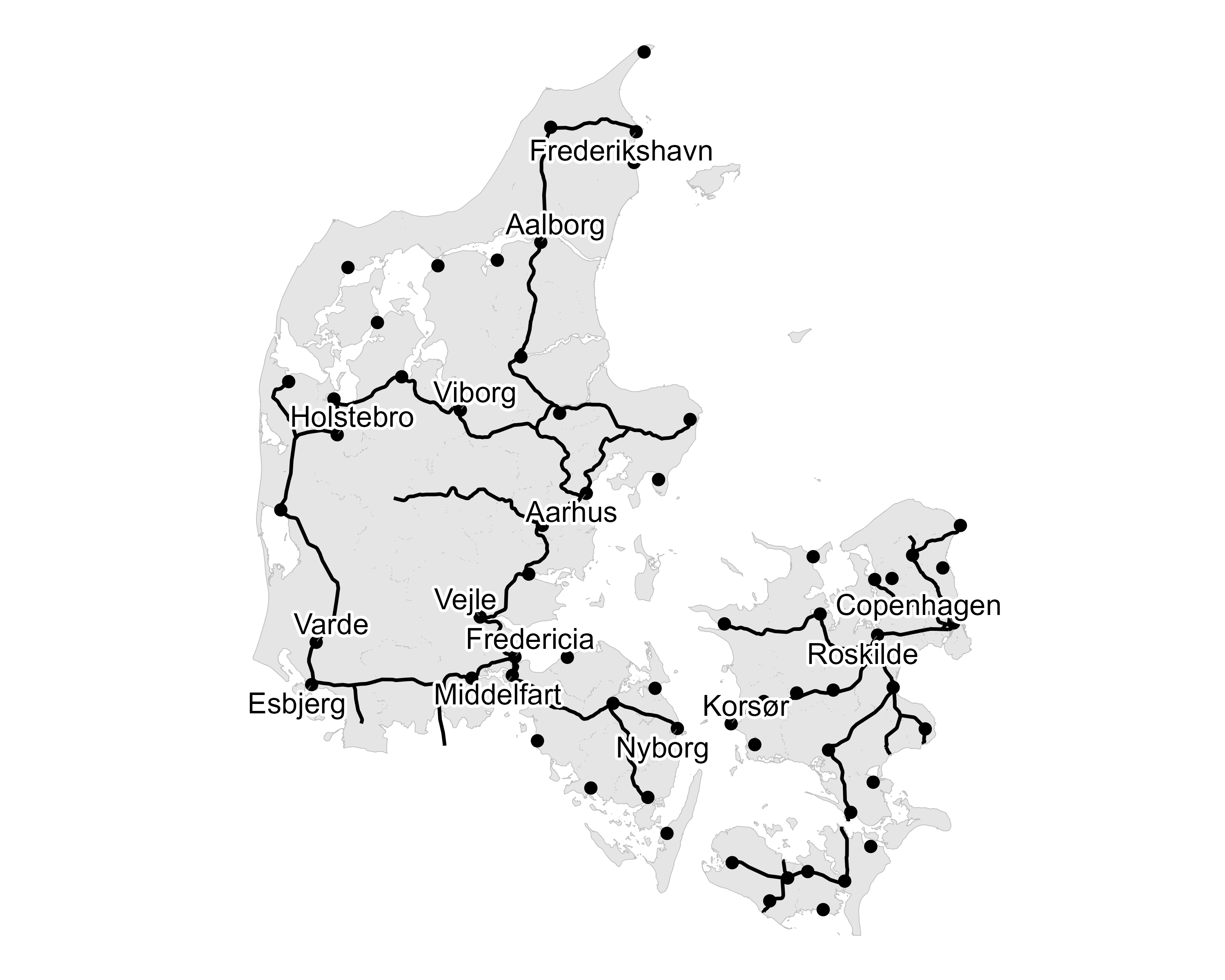} 
        \caption*{(C) 1880}
    \end{minipage}
    \hspace{2cm}
    \begin{minipage}[b]{0.4\textwidth}
        \centering
        \includegraphics[width=\linewidth, trim={5cm 1cm 4.6cm 0.75cm}, clip]{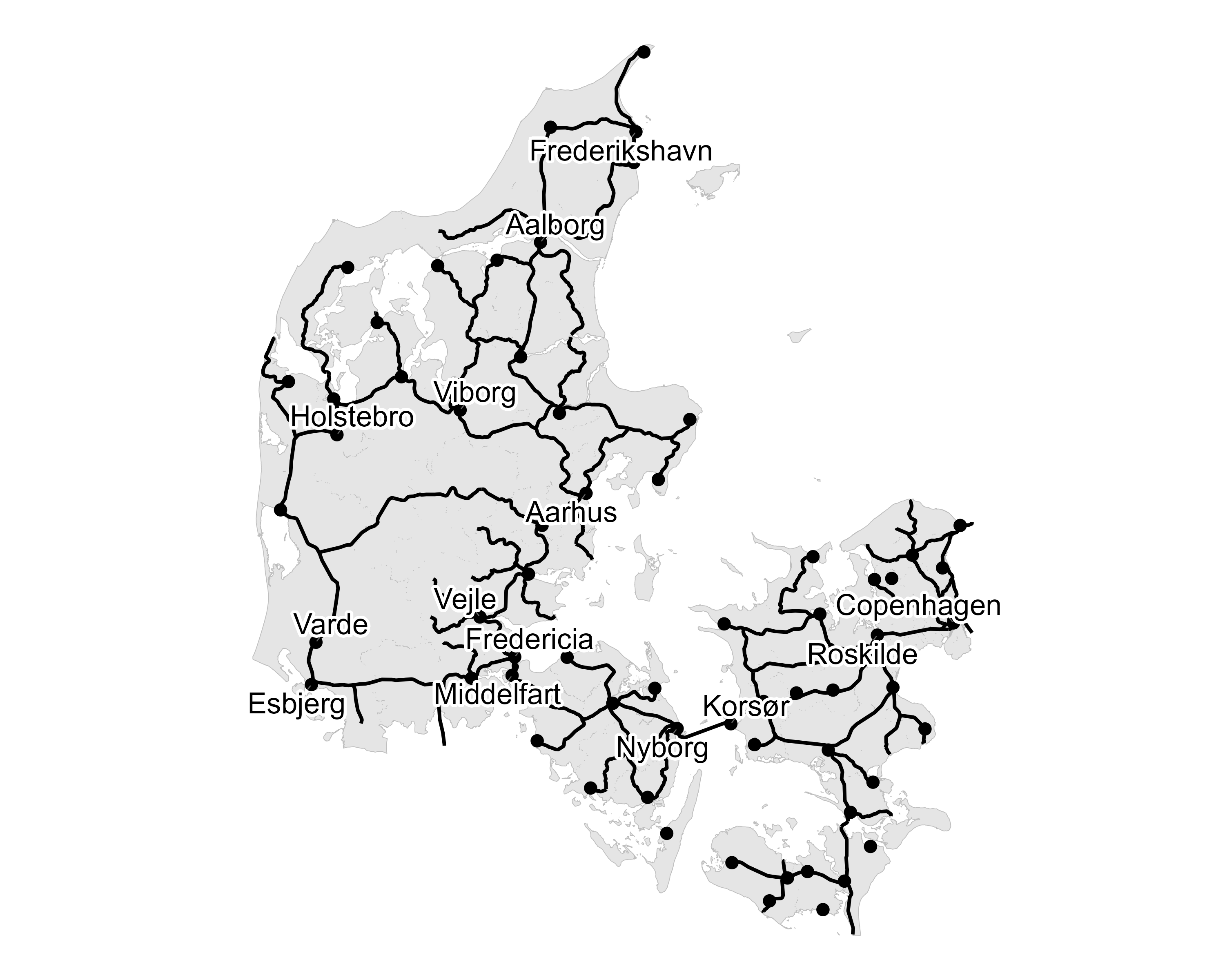} 
        \caption*{(D) 1901}
    \end{minipage}

    \parbox{1\textwidth}{%
        \caption*{\small{\textit{Notes}: This figure displays the Danish railroad network, based on data from \citet{fertner2013}, as solid black lines for the years 1850 (Panel A), 1860 (Panel B), 1880 (Panel C) and 1901 (Panel D). Important nodes that would eventually be connected are shown as black dots. The administrative boundaries of Denmark are taken from \url{https://dataforsyningen.dk/}. For illustrative purposes, the small island of Bornholm is excluded.}}
    }%
    \label{fig:maps_rail}
\end{figure}

Overall, the construction of the railroad network in Denmark was a gradual process, reflecting a combination of economic imperatives and political considerations. Figure \ref{fig:maps_rail} illustrates its expansion from 1850 to 1901, corresponding to the census years used in the analysis of development outcomes. Important nodes\footnote{Defined as the cities holding market town privileges in 1847, the year railroad expansion began, together with Esbjerg and Struer on the west coast, which were strategically targeted as terminals to facilitate trade with England.} are also shown. One important qualification is that although the railroad reached Korsør by 1856, a direct railroad ferry connection across the Great Belt was not established until 1883, increasing transport costs and travel times between Zealand and Funen before then.

\FloatBarrier
\section{Data and Empirical Strategy} \label{sec:empirical_strategy}
We have two primary empirical objectives. First, we aim to estimate the effects of the introduction of the railroad on local economic outcomes. Second, we seek to assess whether railroad expansion facilitated the diffusion of Grundtvigian institutions. We do not attempt to establish a causal hierarchy between these dimensions or to decompose one into the other. Rather, our approach estimates the reduced-form effect of railroad connection on each set of outcomes in parallel, providing a quantitative measure of the extent to which railroads contributed to the joint pattern of economic change and institutional diffusion associated with modern Denmark.

To achieve this, we have matched railroad lines from \citet{fertner2013} to parishes from \citet{digdag}, allowing us to construct a panel dataset that records whether a parish was connected to the railroad in any given year. The resulting dataset covers 1,589 parishes across Denmark, providing near-complete national coverage. Economic outcomes are measured using census data from \citet{mathiesen2022linklives} (1850, 1860, 1880, 1901 censuses), from which we extract parish-level population counts and calculate the child-women ratio (children aged 1 to 5 per woman aged 15 to 45). Additionally, we compute a measure of migration based on the number of individuals residing in a different county from their birthplace. Occupational data is standardized using \citet{dahl2024hisco}, which converts occupational descriptions into HISCO codes.\footnote{We tested 200 randomly sampled observations and report a 95 percent accuracy rate.} From this, we compute parish-level average HISCAM scores and construct measures of the sectoral composition of the labor force. For the latter, we sum all individuals classified into any HISCO major group (0–9) to obtain the total labor force. The manufacturing share is defined as the proportion of workers whose occupations fall in major groups 7, 8, and 9. These groups cover production and related workers, transport equipment operators, and laborers. The non-agricultural share is defined as the proportion of workers whose occupations fall outside major group 6 (farmers), that is, the total labor force minus agricultural workers divided by the total labor force.

To measure the spread of Grundtvigianism, we follow \citet{bentzen2023holy}, using data on the construction of community houses and folk high schools, originally sourced from \cite{trap1906kongeriget} and \cite{trap1928kongeriget}. We employ two types of outcome variables: a dummy variable indicating the presence of a community house or folk high school in a given parish, and an alternative measure that captures exposure through establishments in neighboring parishes. The presence measure captures local siting, while the exposure measure captures spatial spillovers consistent with diffusion beyond parish boundaries.

\begin{figure}[ht]
    \caption{The distribution of Grundtvigianism over time}
    \centering
    \includegraphics[width=0.9\linewidth]{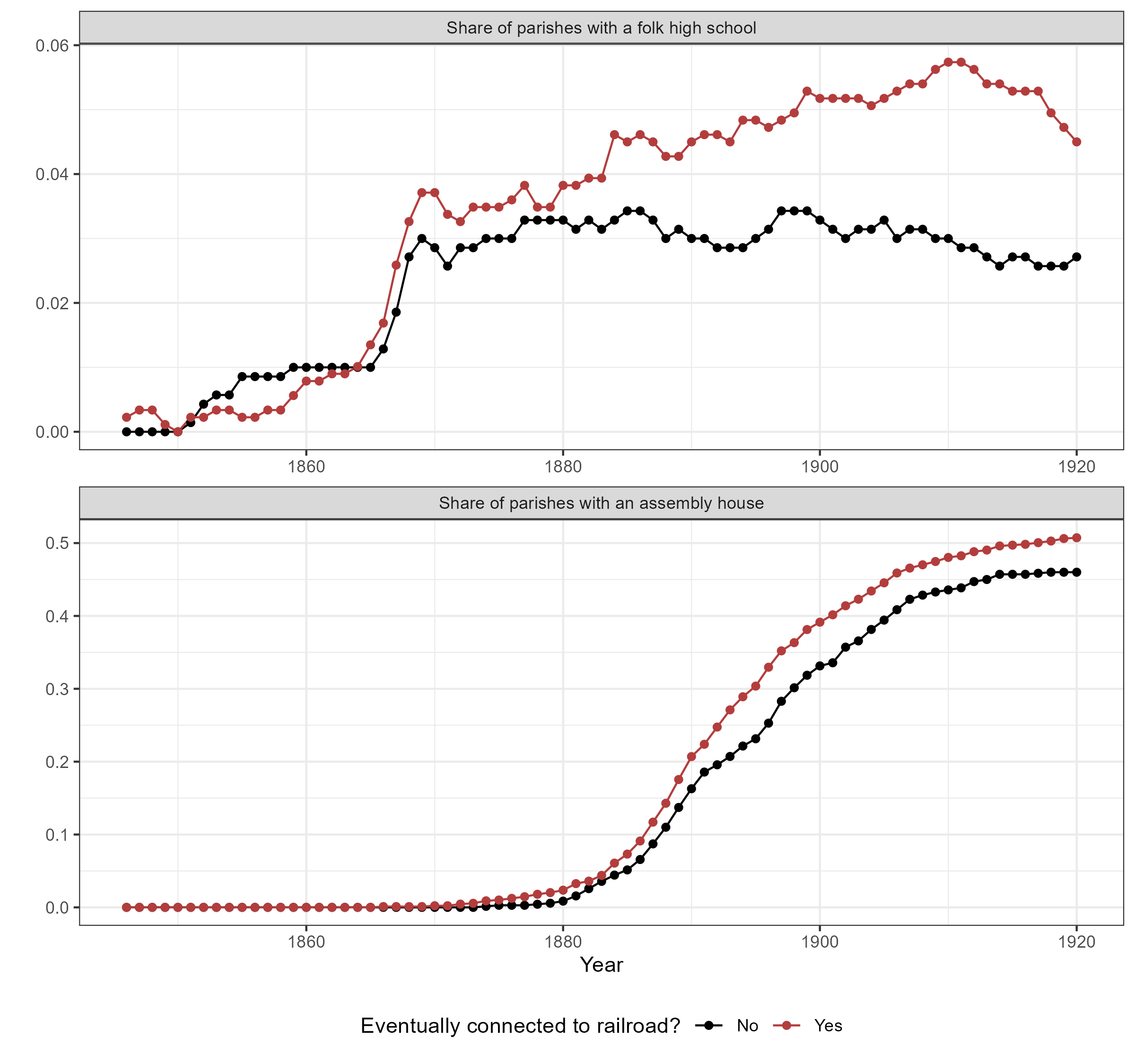}
    \parbox{1\textwidth}{\caption*{\small{\textit{Notes}: This illustrates the spread of Grundtvigianism over time in the form of folk high schools and community houses.}}}
    \label{fig:grundtvig_over_time}
\end{figure}

For the binary outcomes, we compute the share of parishes with community houses and folk high schools over time by railroad status. This is illustrated in Figure \ref{fig:grundtvig_over_time}. Places that were connected to the railroad diverged in terms of community houses and folk high schools, both associated with Grundtvigianism. The gap is considerably larger for folk high schools than for community houses. In 1920, 4.5 percent of connected parishes hosted a folk high school, compared with only 2.5 percent of parishes that were never connected. By contrast, community houses were nearly equally common in both groups, approximately 50 percent versus 45 percent, indicating only a modest difference. A plausible explanation for this pattern is that folk high schools drew students from a wider catchment area, making locations with railroad access more attractive for their establishment, whereas community houses primarily served local rural populations and were therefore less dependent on regional accessibility. More generally, the establishment of both folk high schools and community houses occurred predominantly toward the end of the nineteenth century, when most of the railroad network had already been completed. This timing is consistent with the view that these institutions spread along an existing transport network rather than shaping its expansion.

The local density of Grundtvigian institutions is calculated using an inverse distance-weighted measure of market access, following \citet{Harris1954} and \citet{bentzen2023holy}. This approach assigns greater influence to nearby institutions while still accounting for the presence of more distant ones. Specifically, for each parish $i$, local density is measured as the market access given by

\begin{equation}
    \textit{MA}_i=\sum_{\forall x} dist(i, x)^{-1},
\end{equation}

where $dist(i, x)$ represents the great-circle distance from parish $i$ to each institution $x$. A higher value of this measure indicates a greater local concentration of community houses or folk high schools, suggesting stronger regional diffusion of Grundtvigianism. Conversely, a lower value reflects institutional sparsity. This measure captures spatial spillovers, as the presence of institutions in neighboring parishes contributes to the local density, reflecting the extent to which Grundtvigian influence transcends parish boundaries. 

Table \ref{tbl:summary} presents summary statistics for our main variables.

\begin{table}[!h]
\centering
\caption{Summary statistics}
\centering
\begin{tabular}[t]{lccccc}
\toprule
Variable & N & Mean & SD & Min & Max\\
\midrule
\addlinespace[0.3em]
\multicolumn{6}{l}{\textbf{A. Census}}\\
\hspace{1em}Population & 6356 & 793.777 & 590.528 & 37.000 & 13087.000\\
\hspace{1em}Manufacturing & 6346 & 0.256 & 0.080 & 0.000 & 0.718\\
\hspace{1em}Not agriculture & 6346 & 0.350 & 0.112 & 0.000 & 0.888\\
\hspace{1em}Child-women ratio & 6336 & 0.488 & 0.111 & 0.000 & 1.364\\
\hspace{1em}HISCAM avg & 6355 & 48.631 & 1.503 & 44.235 & 82.850\\
\hspace{1em}Migration & 6356 & 263.233 & 390.587 & 0.000 & 4299.000\\
\hspace{1em}Connected railroad & 6356 & 0.176 & 0.381 & 0.000 & 1.000\\
\addlinespace[0.3em]
\multicolumn{6}{l}{\textbf{B. Grundtvig}}\\
\hspace{1em}Community house & 119175 & 0.171 & 0.377 & 0.000 & 1.000\\
\hspace{1em}Folk high school & 119175 & 0.029 & 0.169 & 0.000 & 1.000\\
\hspace{1em}Density Assembly houses (MA) & 119175 & 4.464 & 5.404 & 0.000 & 20.353\\
\hspace{1em}Density Folk high schools (MA) & 119175 & 0.702 & 0.426 & 0.005 & 3.784\\
\hspace{1em}Connected railroad & 119175 & 0.258 & 0.437 & 0.000 & 1.000\\
\bottomrule
\end{tabular}
\parbox{1\textwidth}{\caption*{\small{\textit{Notes}: Summary statistics for our main variables. Panel A contains census data from \cite{mathiesen2022linklives}, Panel B contains data on Grundtvigianism in the form of community houses and folk high schools from \cite{bentzen2023holy}.}}} \label{tbl:summary}
\end{table}

Our baseline empirical strategy employs a two-way fixed effects (TWFE) estimator:

\begin{equation}
    y_{it} = \alpha_i + \alpha_t + \textit{connected railroad}_{it} \beta + X_{it}\gamma + \varepsilon_{it},
\end{equation}

where $y_{it}$ represents the outcome of interest, $\alpha_i$ and $\alpha_t$ denote parish and year fixed effects, respectively, and $\textit{connected railroad}_{it}$ is an indicator variable for whether a parish has a railroad passing through it in a given year. $X_{it}$ denotes the vector of controls and $\gamma$ denotes a vector of associated parameters. The controls include distance to Hamburg, distance to Copenhagen, parish population in 1801, county-year fixed effects, and distance to the Ox Road, a major historical transport route.\footnote{For the TWFE specifications, these covariates enter in discretized form, where continuous measures are divided into deciles and interacted with the year to allow for flexible time-varying controls, whereas in the \citet{Callaway2021} estimator the corresponding time-invariant versions of these variables are used.} The distance measures capture the mechanically distinctive geography of the earliest connected parishes, which are located close to Copenhagen and Hamburg but relatively distant from older overland corridors such as the Ox Road, while parish population in 1801 and county-year fixed effects control for pre-railroad settlement size and time-varying regional shocks that may jointly affect railroad placement and local outcomes. The coefficient $\beta$ captures the average effect of railroad connection on $y_{it}$.

A key concern with this specification is the presence of heterogeneous treatment effects and staggered adoption, which can lead to problematic comparisons where already treated units are used as controls \citep{GoodmanBacon2021}. To address this, we follow the approach of \cite{Callaway2021}, which estimates a series of $2\times2$ classical difference-in-differences models, allowing for flexible aggregation to identify group-specific effects. Specifically, we use parishes that never receive a railroad connection as the control group. This is because our Grundtvigian outcomes are measured through 1920, while the railroad network continued to expand only until 1929. By 1920, only a small number of parishes remain ``not-yet-treated'', and these late connections are minor branch lines, making them an unsuitable comparison group.

\begin{figure}[ht]
    \caption{Distribution of variables in 1850}
    \centering
    \includegraphics[width=1\linewidth]{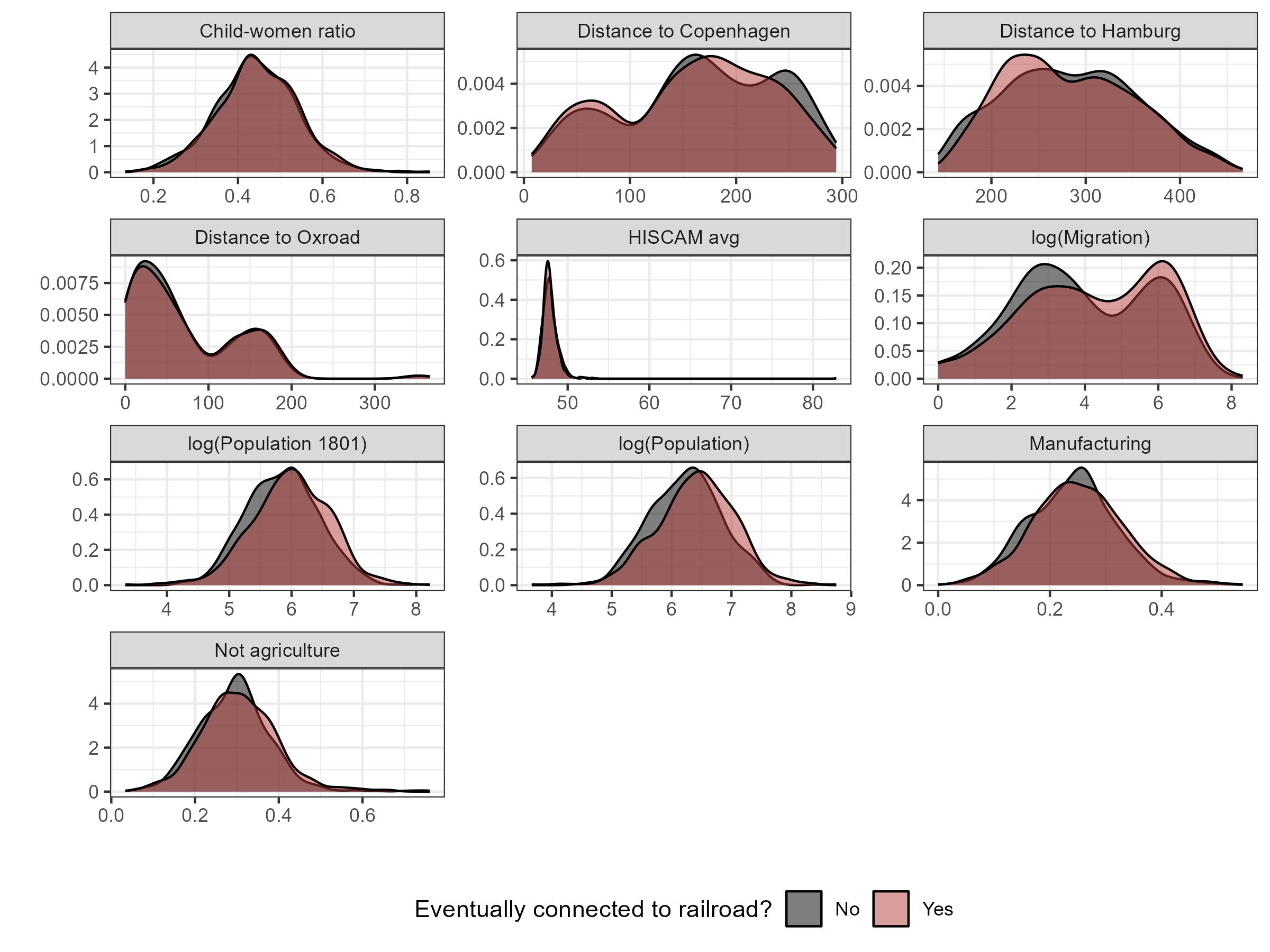}
    \parbox{1\textwidth}{\caption*{\small{\textit{Notes}: Density estimates of a number of variables from the census data and other covariates by railroad status. Eight parishes (of 1589), which were already connected in 1850, were excluded from these plots. }}}\label{fig:densities}
\end{figure}

Another concern is selection into railroad connection: connected parishes may have differed systematically from never-connected ones, making the latter a poor comparison group. To assess how comparable the two groups are before treatment, Figure \ref{fig:densities} shows density estimates for the main census outcomes in 1850 (before any railroad connection), as well as a number of long-run growth-related covariates. The figure shows substantial overlap in the distributions between parishes that eventually gained railroad access and those that did not, indicating clear common support in key pre-treatment characteristics.\footnote{Appendix~\ref{appendix:balance} reports balance and distribution tests. Following \citet{Baker_et_al_2025}, we view these checks (together with what is presented here in the paper) as indirect evidence: they do not establish parallel trends but do demonstrate overlap (common support), a prerequisite for credible difference-in-differences.} The balance tests in Appendix~\ref{appendix:balance} document some imbalance in levels. Parishes that were eventually connected were, on average, more populous in 1850, received more internal migrants, and had higher manufacturing and non-agricultural employment shares.
These imbalances are not by themselves problematic for the identification.\footnote{See assumption 6 in \citet{Callaway2021}. Figure~\ref{fig:densities_by_treat_year} in Appendix~\ref{densities_by_year} supplements this with density plots, which are grouped by when the parish was connected. These show that the differences are most pronounced for the small cohort connected by 1860, whose location along the Copenhagen to Korsør line makes its geography mechanically distinctive.} As a further check, Appendix~\ref{appendix:excl_early} re-estimates the baseline specifications excluding the parishes connected by 1860, which drive most of the imbalance; the results are similar to the baseline estimates.

\begin{figure}[ht]
    \caption{Census outcomes over time}
    \centering
    \includegraphics[width=1.0\linewidth]{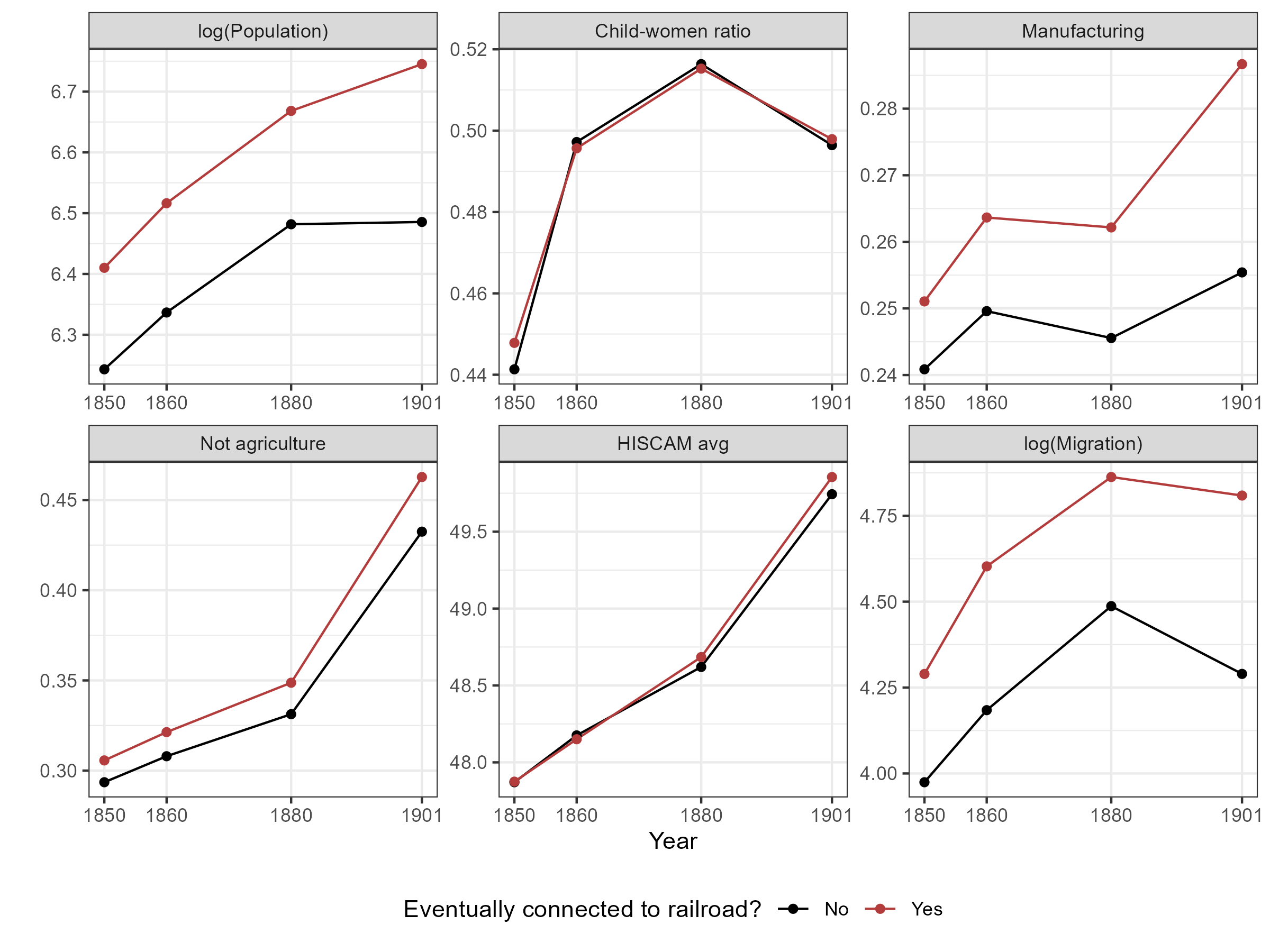}
    \parbox{1\textwidth}{\caption*{\small{\textit{Notes}: Unconditional parish-level means of the six census outcomes in each census year (1850, 1860, 1880, 1901), separately for parishes eventually connected to the railroad and parishes never connected. Eight parishes (of 1589), which were already connected in 1850, were excluded from these plots.}}}
    \label{fig:census_outcomes_over_time}
\end{figure}

Figure~\ref{fig:census_outcomes_over_time} complements this comparison by plotting unconditional means of the six census outcomes in each census year, separately for parishes that were eventually connected and parishes that were never connected. Three patterns stand out. First, the two groups track each other closely for the child-women ratio and average HISCAM throughout the period, foreshadowing the absence of fertility effects and the modest effects on socio-economic status. Second, parishes that were eventually connected start from higher levels in the remaining variables, consistent with the balance tests reported in Appendix~\ref{appendix:balance}. Third, the two groups evolve broadly in parallel over the early decades, and the visible divergence in population, migration, manufacturing, and the non-agricultural share is concentrated after 1880, when the main trunk lines were in place. Because connection is staggered across the period, these raw means mix pre- and post-connection observations within the treated group, so we read the figure as descriptive. The event-study estimates below provide an assessment of pre-trends.

To further assess the plausibility of parallel trends and better understand treatment-effect dynamics, we additionally show average treatment effects at different lengths of exposure in event-study plots. The resulting coefficients trace out the evolution of outcomes before and after connection. We report these estimates for the census-based outcomes in Figure~\ref{fig:decompose_census_dynamic} and for the Grundtvigian outcomes in Figure~\ref{fig:decompose_grundtvig_dynamic}.

\FloatBarrier
\section{Results}

\subsection{Railroads and Local Development}
Table \ref{tbl:rail_local_dev} reports the estimated effects of railroad connection on our main economic outcomes. Panel A presents Two-Way Fixed Effects (TWFE) estimates, while Panel B reports the \citet{Callaway2021} estimates. Column (1) shows that railroad access is associated with an increase in population of about 7 percent. Column (2) indicates no detectable effect on fertility, measured by the child-women ratio. Columns (3) and (4) show positive effects on structural change: connected parishes exhibit, on average, a 1.8-percentage-point higher manufacturing employment share and a 2-percentage-point higher non-agricultural employment share, conditional on controls. Column (5) suggests a small increase in average HISCAM, but the estimate is only marginally significant. Finally, Column (6) indicates higher internal migration, proxied by the number of residents born in a different county: railroad access is associated with 10 percent more residents born outside the county. Taken together, these patterns suggest that the population response reflects, at least in part, a reallocation of settlement and economic activity toward connected locations, consistent with evidence from Sweden, England and Wales, and Switzerland \citep{berger_enflo_2017,bogart2022,Büchel_and_Kyburz_2018}.

Standard errors are clustered at the parish level in the baseline specification, but because treatment is spatially correlated, we also report in Appendix~\ref{Alt_ses_census_and_grundtvig} results with standard errors clustered at the wider county level and, for the Two-Way Fixed Effects (TWFE) specifications, heteroskedasticity- and spatial-autocorrelation-consistent (Conley) standard errors at different cutoffs (10 km, 25 km, and 50 km).\footnote{Conley standard errors are not readily available for the \citet{Callaway2021} estimator. For those specifications, clustering at the county level is the most conservative adjustment we implement.} The results remain qualitatively similar.

\begin{landscape}
    \begin{table}[ht]
    \caption{Railroads and Local Development}
    \centering
    \begin{tabular}{lcccccc}
  \toprule
  Outcome: & log(Pop.) & Child-women ratio & Manufacturing & Not Agriculture & HISCAM avg & log(Migration) \\
           & (1) & (2) & (3) & (4) & (5) & (6) \\
  \midrule
  \multicolumn{7}{l}{\textbf{A. TWFE estimates}}\\
  Connected railroad &  0.0664$^{***}$ & 0.0016$^{}$ & 0.0173$^{***}$ & 0.0186$^{***}$ & 0.0862$^{}$ & 0.1353$^{***}$  \\
                    &  (0.0094) & (0.0047) & (0.0032) & (0.0039) & (0.0641) & (0.0397)  \\
  \cmidrule(lr){2-7}
  Observations      &  6356 & 6336 & 6346 & 6346 & 6355 & 6264  \\
  Mean of outcome   &  6.4738 & 0.4883 & 0.2559 & 0.3496 & 48.6308 & 4.4198  \\
  \midrule
  \multicolumn{7}{l}{\textbf{B. Callaway and Sant'Anna estimates}}\\
  Connected railroad &  0.0703$^{***}$ & 0.0050$^{}$ & 0.0177$^{***}$ & 0.0203$^{***}$ & 0.1252$^{*}$ & 0.0998$^{**}$  \\
                    &  (0.0099) & (0.0050) & (0.0035) & (0.0046) & (0.0741) & (0.0463)  \\
  \cmidrule(lr){2-7}
  Observations      &  6320 & 6240 & 6280 & 6280 & 6316 & 6000  \\
  Mean of outcome   &  6.4716 & 0.4884 & 0.2555 & 0.3489 & 48.6185 & 4.4884  \\
  \bottomrule
\end{tabular}

    \parbox{1.2\textwidth}{\caption*{\small{\textit{Notes}: This table reports estimates of the effect of railroad connection on parish-level economic outcomes. Panel A shows Two-Way Fixed Effects (TWFE) estimates, and Panel B shows estimates from \citet{Callaway2021}. Columns (1) through (6) report population, fertility (child-women ratio), the manufacturing employment share, the non-agricultural employment share, average HISCAM, and internal migration. All specifications include the covariates as described in Section 5. Standard errors are clustered at the parish level. *** $p< 0.01$ ** $p< 0.05$ * $p< 0.10$.}}}\label{tbl:rail_local_dev}
    \end{table}
\end{landscape}

\begin{figure}[ht]
    \caption{Dynamic Decomposition (census)}
    \centering
    \begin{minipage}[b]{0.3\textwidth}
        \centering
        \includegraphics[width=\textwidth]{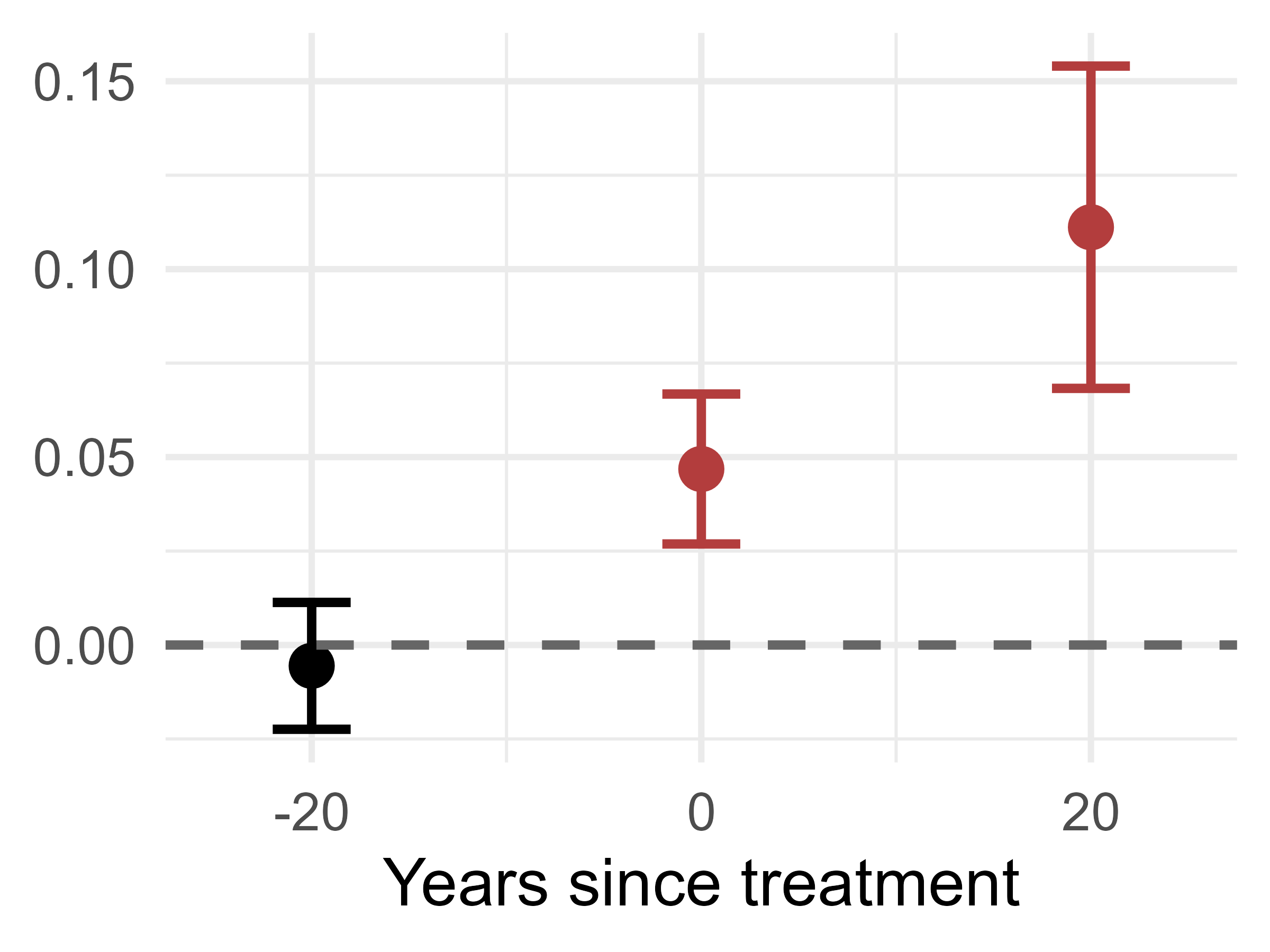}
        \caption*{(A) log(Pop.)}
    \end{minipage}
    \hfill
    \begin{minipage}[b]{0.3\textwidth}
        \centering
        \includegraphics[width=\textwidth]{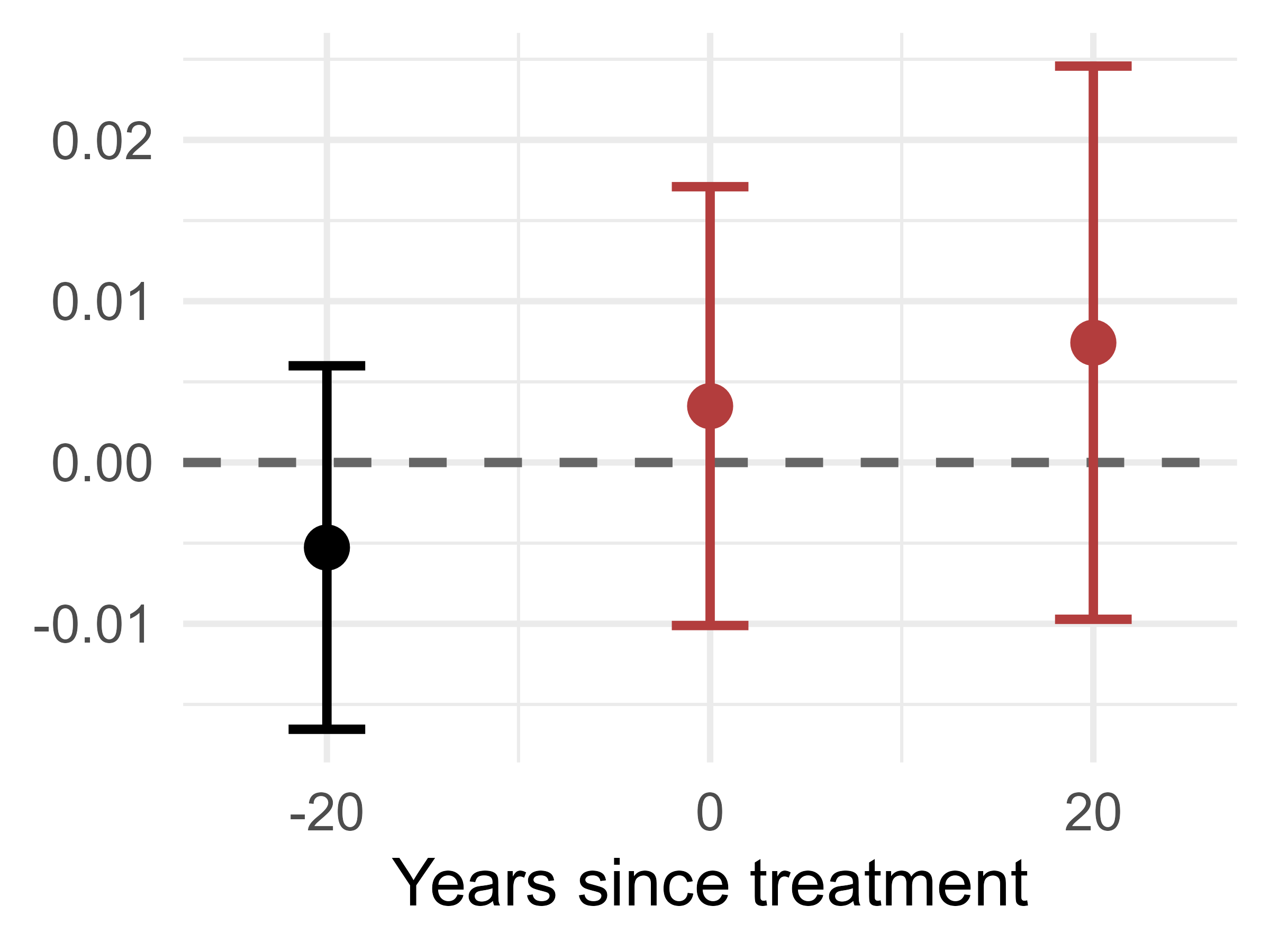}
        \caption*{(B) Child-women ratio}
    \end{minipage}
    \hfill
    \begin{minipage}[b]{0.3\textwidth}
        \centering
        \includegraphics[width=\textwidth]{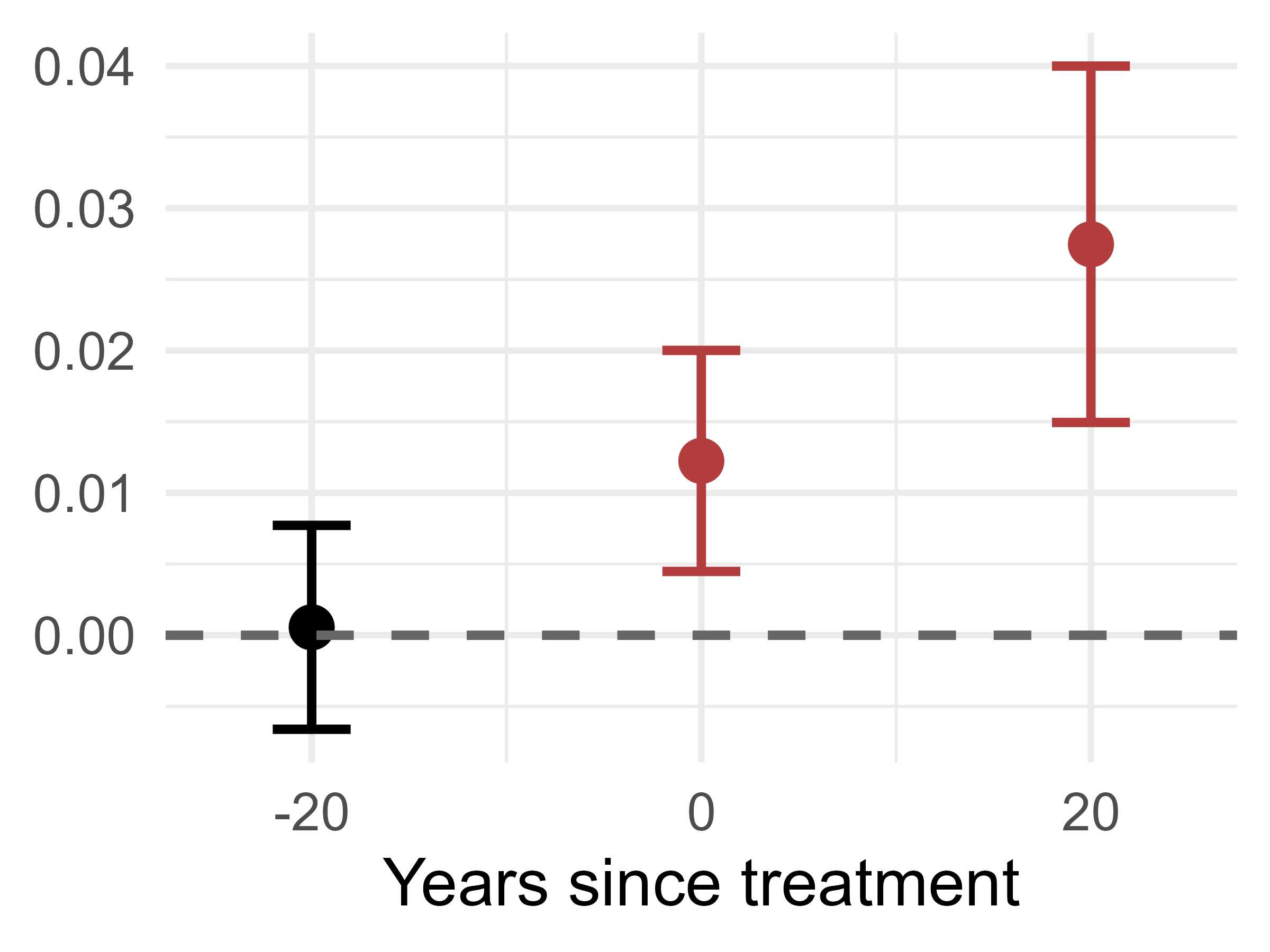}
        \caption*{(C) Manufacturing}
    \end{minipage}
    
    \vspace{1em}

    \begin{minipage}[b]{0.3\textwidth}
        \centering
        \includegraphics[width=\textwidth]{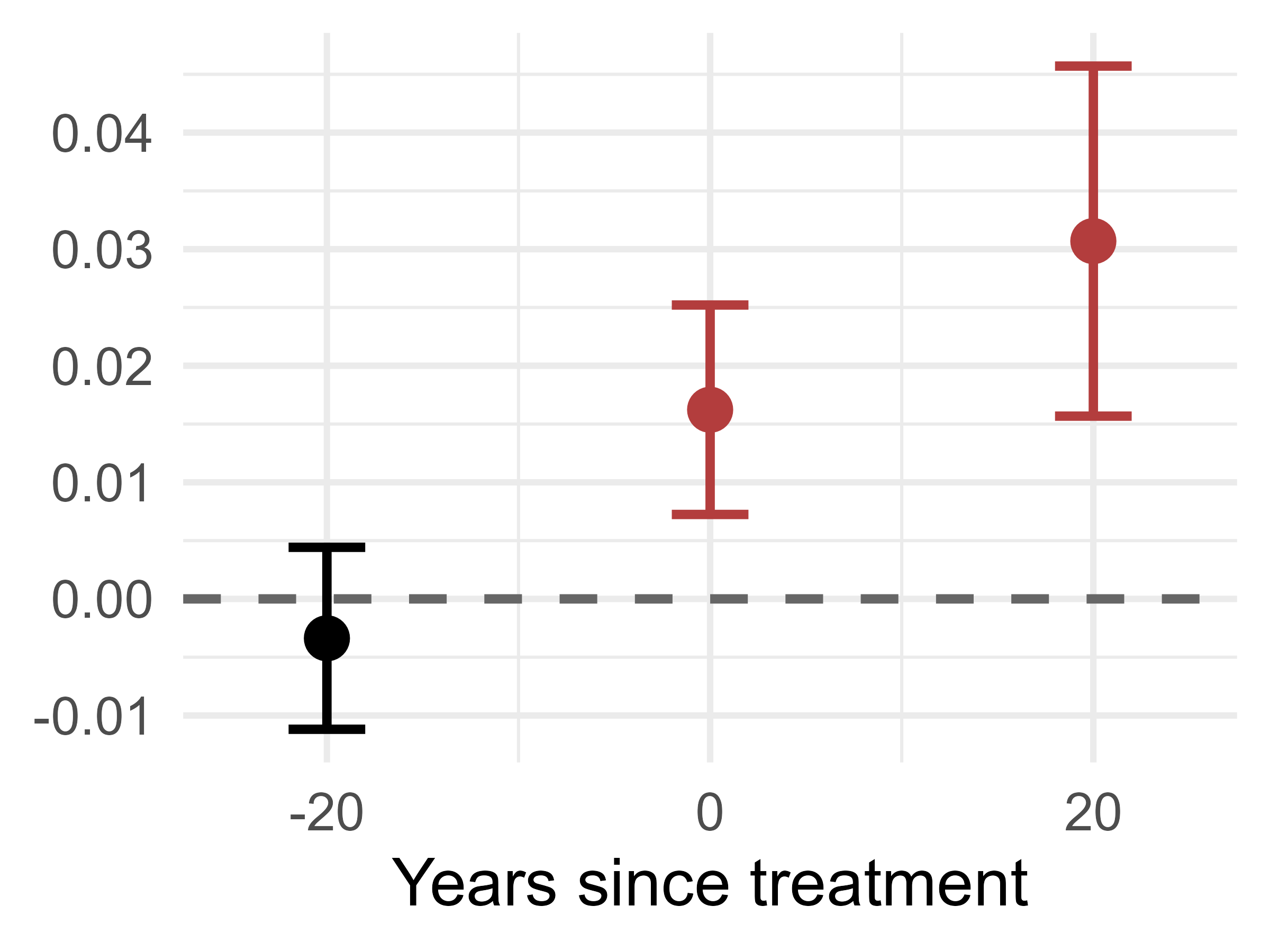}
        \caption*{(D) Non-agricultural}
    \end{minipage}
    \hfill
    \begin{minipage}[b]{0.3\textwidth}
        \centering
        \includegraphics[width=\textwidth]{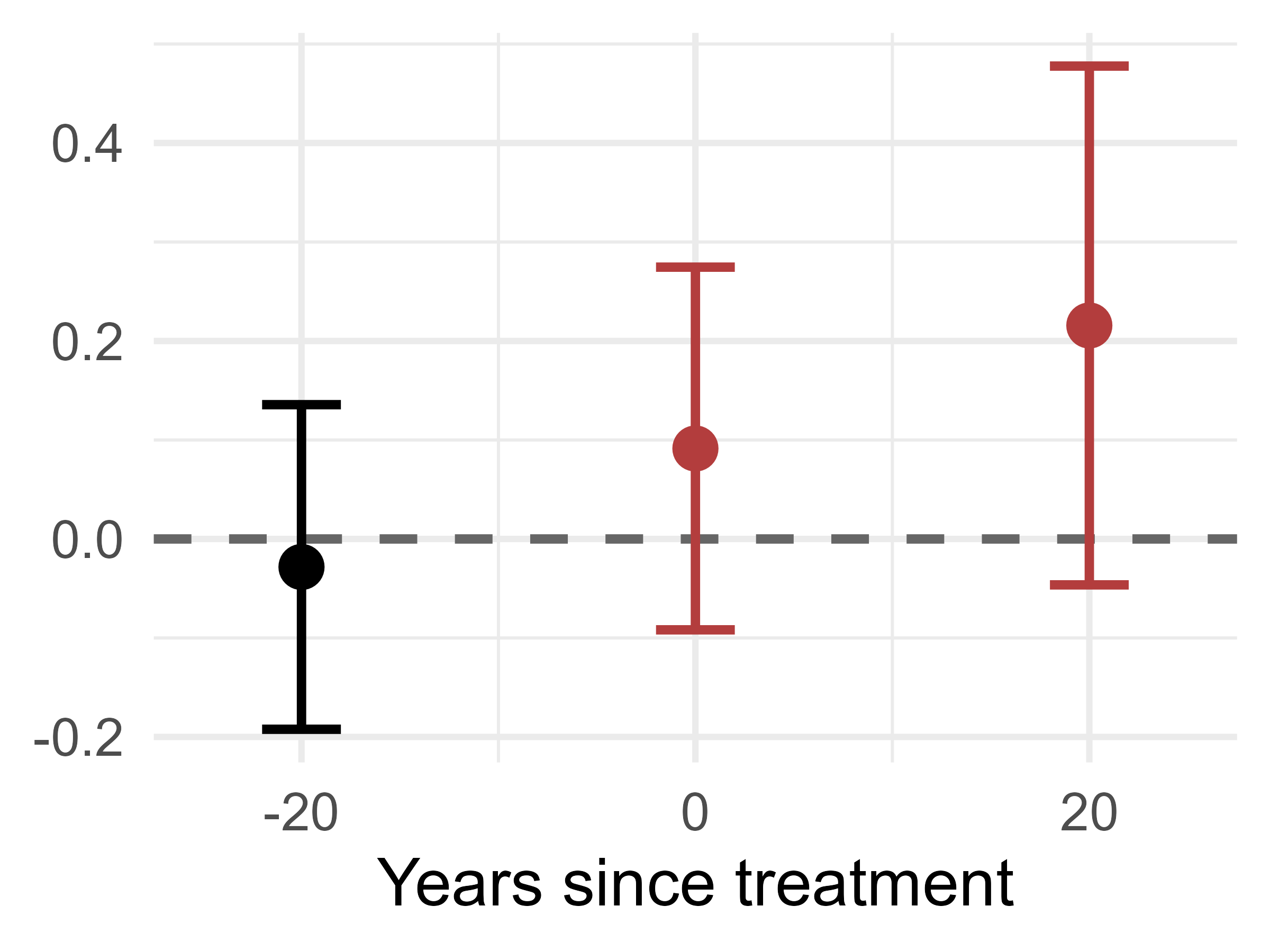}
        \caption*{(E) HISCAM}
    \end{minipage}
    \hfill
    \begin{minipage}[b]{0.3\textwidth}
        \centering
        \includegraphics[width=\textwidth]{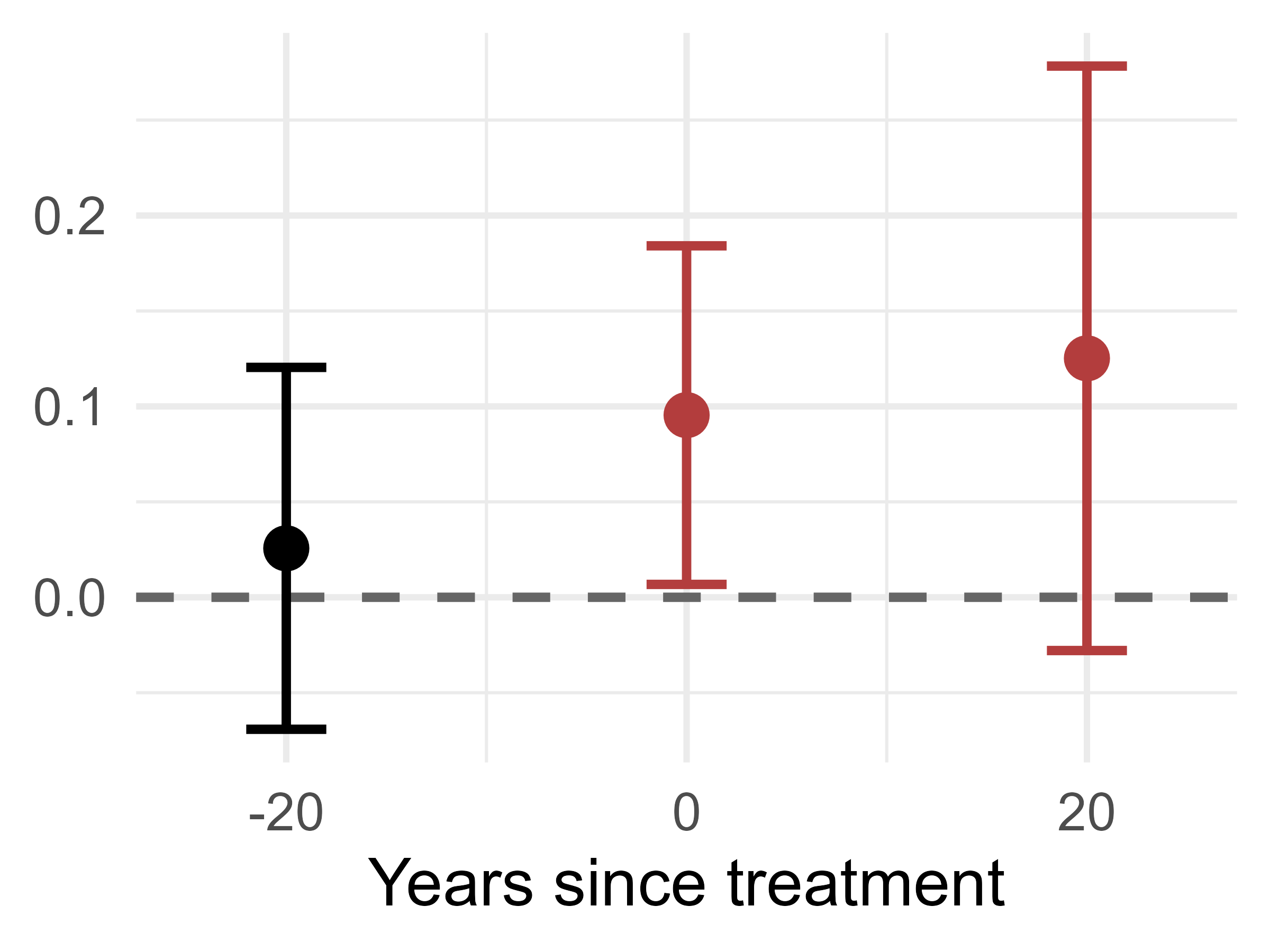}
        \caption*{(F) log(Migration)}
    \end{minipage}

    \parbox{1\textwidth}{%
        \caption*{\small{\textit{Notes}: Event-study estimates by years since railroad connection for six census-based outcomes, using the \citet{Callaway2021} estimator relative to never-treated parishes and conditioning on the covariates described in Section~\ref{sec:empirical_strategy}. The event study is restricted to event times between -20 and +20 years relative to treatment. Whiskers indicate 95\% confidence intervals.}}
    }\label{fig:decompose_census_dynamic}
\end{figure}

To gain insight into dynamics and to assess the plausibility of parallel trends, Figure \ref{fig:decompose_census_dynamic} shows how the effects of railroad connection on our census outcomes vary with years of exposure. We restrict attention to the event-time window from -20 to +20 years. This excludes the sparsely identified tails around -40 and +40 years, which rely on only a narrow set of parishes connected very late or very early and are correspondingly imprecise.

The event-study coefficients show no systematic pre-trends. Estimates prior to connection are small and statistically indistinguishable from zero across outcomes. In contrast, clear effects emerge only after connection for several outcomes. This pattern supports the conditional parallel trends assumption underlying our difference-in-differences design for the economic outcomes.

Panel (A) shows that population responds immediately to connection and that the effect is larger after twenty years of exposure, consistent with gradual adjustment as improved market access and mobility attract migrants and economic activity. Panel (B) shows no evidence of fertility responses: coefficients for the child-women ratio remain close to zero across exposure horizons. Panels (C) and (D) show parallel dynamics for structural change. Manufacturing and the non-agricultural employment share increase immediately after connection and grow in the following decades, consistent with a gradual reallocation of labor out of agriculture. Panel (E) shows no clear effect on average HISCAM. Panel (F) shows an immediate increase in migration upon connection, but little evidence that this response grows with additional exposure.

Overall, these results indicate that railroads shaped local development not only through population growth but also through migration and structural change.\footnote{Appendix~\ref{regresssionsnocontrols}, Table~\ref{tbl:rail_census_nocontrol} reports the corresponding regressions without controls. The results remain qualitatively similar; the estimate on average HISCAM becomes more precisely estimated but remains small in magnitude.}

\subsection{Railroads and Grundtvigianism}

Table \ref{tbl:rail_grundtvig} reports estimates of the effect of railroad connection on Grundtvigian institutions. Our hypothesis is that Grundtvigian institutions spread with the railroad, and accordingly we test both whether there is a higher density of Grundtvigian institutions in the area around the railroad and whether connected parishes were more likely to host these institutions themselves.

Columns (1) and (2) use parish-level indicators for the presence of a community house or a folk high school. Columns (3) and (4) use inverse-distance-weighted market access measures of local density, which capture exposure to nearby institutions and thus account for establishments in surrounding parishes. Because both institutions often served residents beyond parish boundaries, local density provides a more appropriate measure of effective reach than the parish-level indicators alone. Conceptually, this maps naturally to a diffusion mechanism: rail connection facilitates exposure to new ideas, and adoption follows.

\begin{table}[ht]
\caption{Railroads and Grundtvigianism}
\centering
\footnotesize
\resizebox{\textwidth}{!}{%
\begin{tabular}{lcccc}
  \toprule
  Outcome: & Community house & Folk high school & \makecell{Density Community \\ houses (MA)} & \makecell{Density Folk High \\ Schools (MA)} \\
           & (1) & (2) & (3) & (4)  \\
  \midrule
  \multicolumn{5}{l}{\textbf{A. TWFE estimates}}\\
  Connected railroad &  0.0357$^{***}$ & 0.0108$^{*}$ & 0.0768$^{***}$ & 0.0125$^{**}$  \\
                    &  (0.0136) & (0.0063) & (0.0269) & (0.0062)  \\
  \cmidrule(lr){2-5}
  Observations      &  119175 & 119175 & 119175 & 119175  \\
  Mean of outcome   &  0.1715 & 0.0295 & 4.4639 & 0.7024  \\
  \midrule
  \multicolumn{5}{l}{\textbf{B. Callaway and Sant'Anna estimates}}\\
  Connected railroad &  0.0037$^{}$ & 0.0168$^{**}$ & 0.1450$^{***}$ & 0.0188$^{**}$  \\
                    &  (0.0137) & (0.0077) & (0.0238) & (0.0080)  \\
  \cmidrule(lr){2-5}
  Observations      &  119100 & 119100 & 119100 & 119100  \\
  Mean of outcome   &  0.1716 & 0.0295 & 4.4646 & 0.7026  \\
  \bottomrule
\end{tabular}
}

\parbox{0.99\textwidth}{\caption*{\small{\textit{Notes}: This table reports estimates of the effect of railroad connection on Grundtvigian institutions. Panel A shows Two-Way Fixed Effects (TWFE) estimates, and Panel B shows estimates from \citet{Callaway2021}. Columns (1) and (2) report parish-level indicators for community houses and folk high schools. Columns (3) and (4) report inverse-distance-weighted measures of local density. All specifications include the covariates as described in Section~\ref{sec:empirical_strategy}. Standard errors are clustered at the parish level. *** $p< 0.01$ ** $p< 0.05$ * $p< 0.10$.}}}\label{tbl:rail_grundtvig}
\end{table}

Panel A of Table \ref{tbl:rail_grundtvig} presents the Two-Way Fixed Effects (TWFE) estimates, while Panel B reports the Callaway and Sant'Anna estimates. Under TWFE, railroad access is associated with a higher probability that a parish has a community house (Column 1) and, with lower precision, a folk high school (Column 2). By contrast, the Callaway and Sant'Anna estimates indicate a statistically significant increase for folk high schools but no detectable effect for community houses. We focus on the Callaway and Sant'Anna estimates because they account for staggered adoption and treatment-effect heterogeneity. Accordingly, we find no significant effect of railroad access on the probability that a parish hosts a community house, while railroad access increases the probability of hosting a folk high school by about 1.7 percentage points. Taken together, these results suggest that folk high schools were more directly tied to connectivity, whereas community houses were not, consistent with a diffusion mechanism in which railroads facilitate the arrival of ideas without mechanically shifting local siting.

Columns (3) and (4) use inverse-distance-weighted local density as the outcome, which more directly captures exposure through neighboring parishes. These results are consistent with diffusion rather than mechanical siting in connected parishes. Community houses---and, to a lesser extent, folk high schools---did not require rail access to operate once established; what mattered was that the organizational form arrived and could be adopted locally. Under our baseline parish clustering, both the TWFE and Callaway--Sant'Anna estimates indicate positive and statistically significant effects on the density of community houses and folk high schools.\footnote{Appendix~\ref{regresssionsnocontrols}, Table~\ref{tbl:rail_grundtvig_nocontrol} reports the corresponding regressions without controls. The results remain qualitatively similar.} Overall, the evidence indicates that railroad access increased local exposure to Grundtvigian institutions and, at the same time, increased the probability that connected parishes hosted a folk high school.

The Grundtvigian estimates warrant a more cautious reading than the economic results. Because railroad lines run through neighboring parishes and the density measure aggregates information across neighboring parishes, both treatment and outcomes may be spatially correlated and the appropriate level of clustering is not obvious. Appendix~\ref{Alt_ses_census_and_grundtvig} therefore reports standard errors clustered at the county level and, for the TWFE specifications, Conley spatial HAC standard errors at several cutoffs. These adjustments leave the point estimates unchanged but inflate the standard errors, and they do so more for the institutional outcomes than for the economic ones, so that some of the Grundtvigian estimates lose statistical significance under the most conservative adjustments of spatial dependence. Appendix \ref{regresssionsnocontrols}, Table \ref{tbl:rail_grundtvig_nocontrol} further shows some sensitivity to the exclusion of controls. Since identification rests on parallel trends conditional on covariates, the specifications with controls remain our preferred ones, but we note the sensitivity. Taken together, we read the institutional results as suggestive evidence that railroads facilitated the diffusion of Grundtvigian institutions, but with less statistical certainty than for the economic effects.

\begin{figure}[ht]
    \caption{Dynamic Decomposition (Grundtvig)}
    \centering
    \begin{minipage}[b]{0.45\textwidth}
        \centering
        \includegraphics[width=\textwidth]{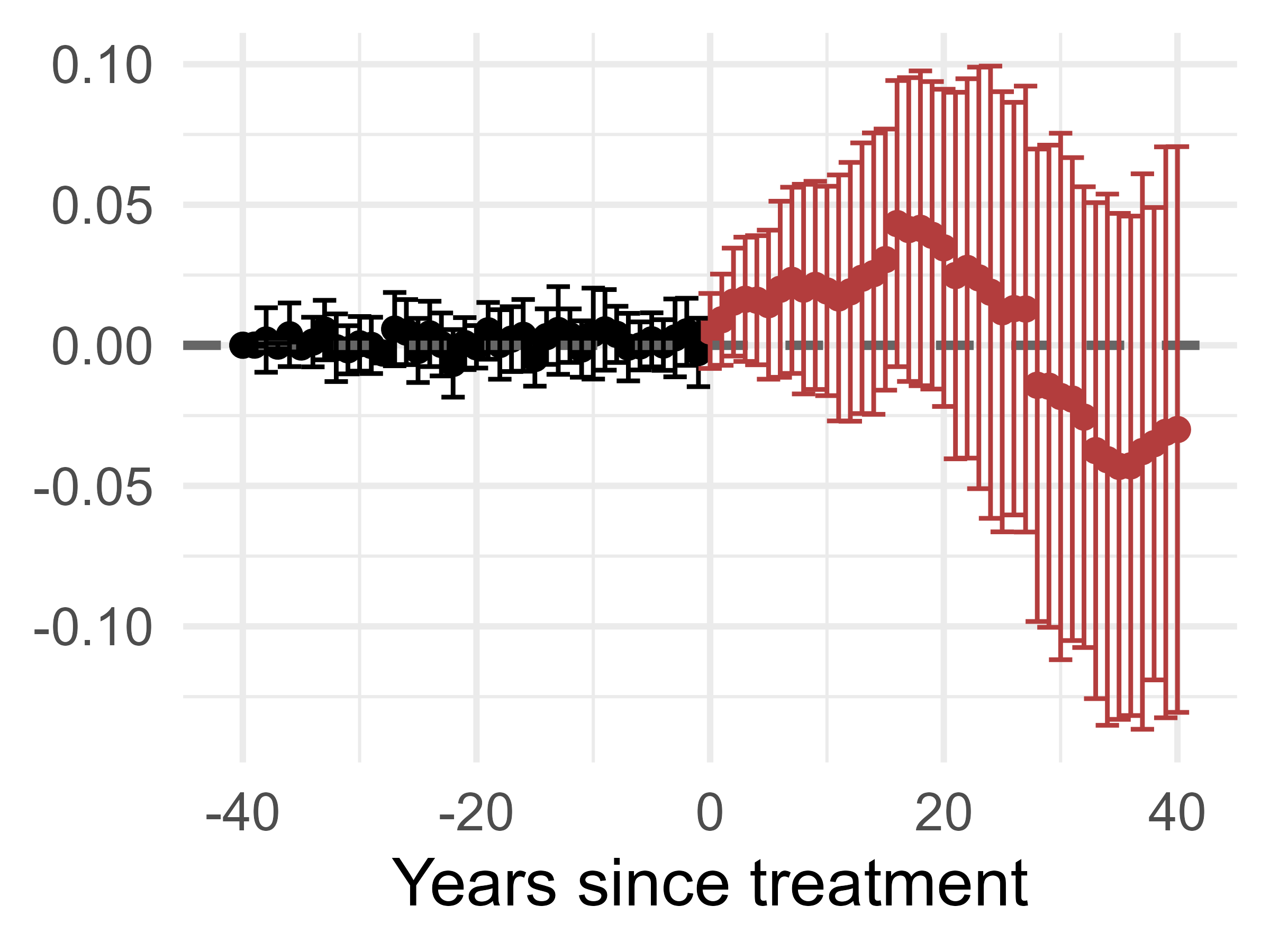}
        \caption*{(A) Community house}
    \end{minipage}
    \hspace{0.5cm}
    \begin{minipage}[b]{0.45\textwidth}
        \centering
        \includegraphics[width=\textwidth]{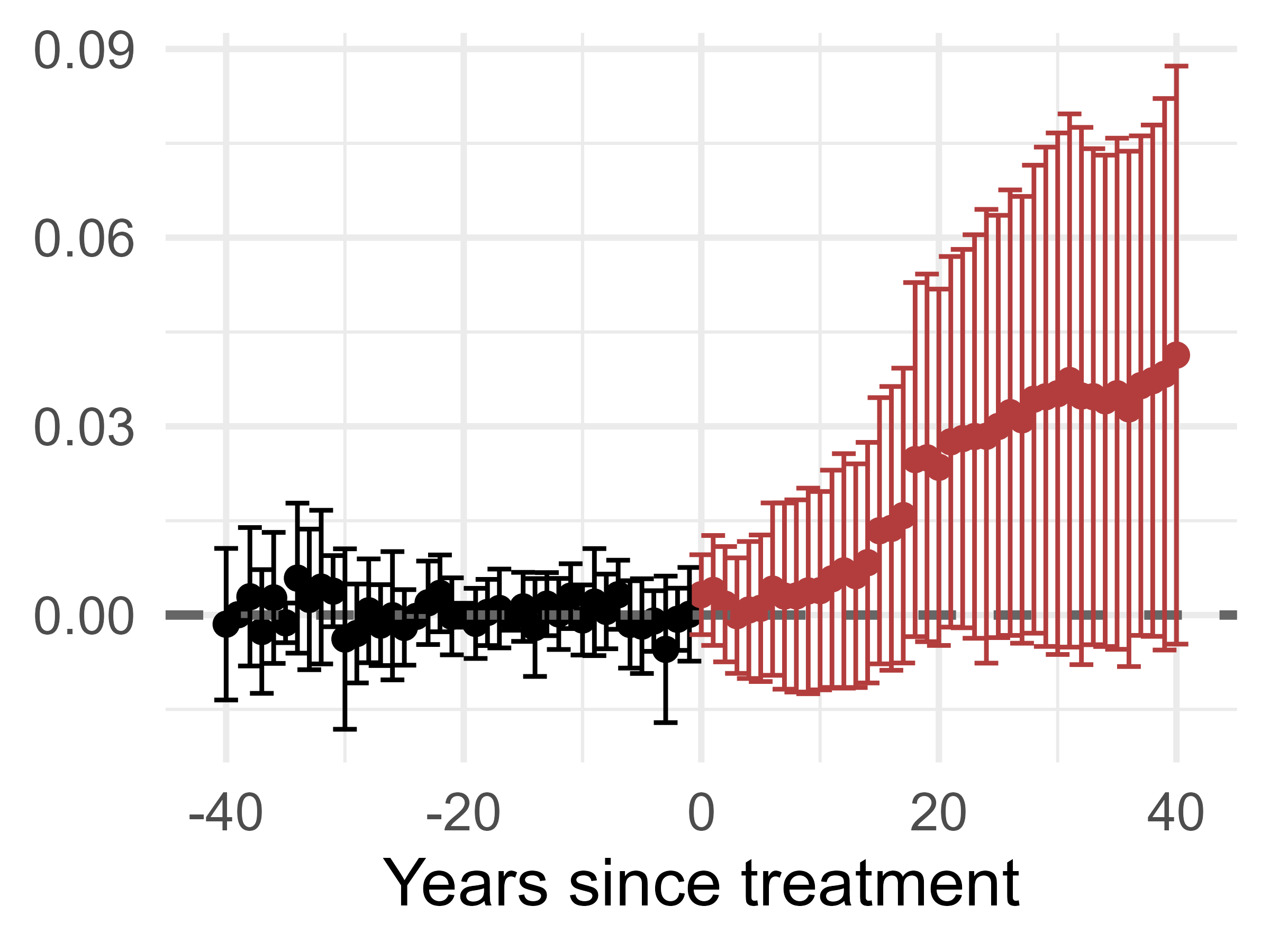}
        \caption*{(B) Folk high school}
    \end{minipage}

    \vspace{1em}

    \begin{minipage}[b]{0.45\textwidth}
        \centering
        \includegraphics[width=\textwidth]{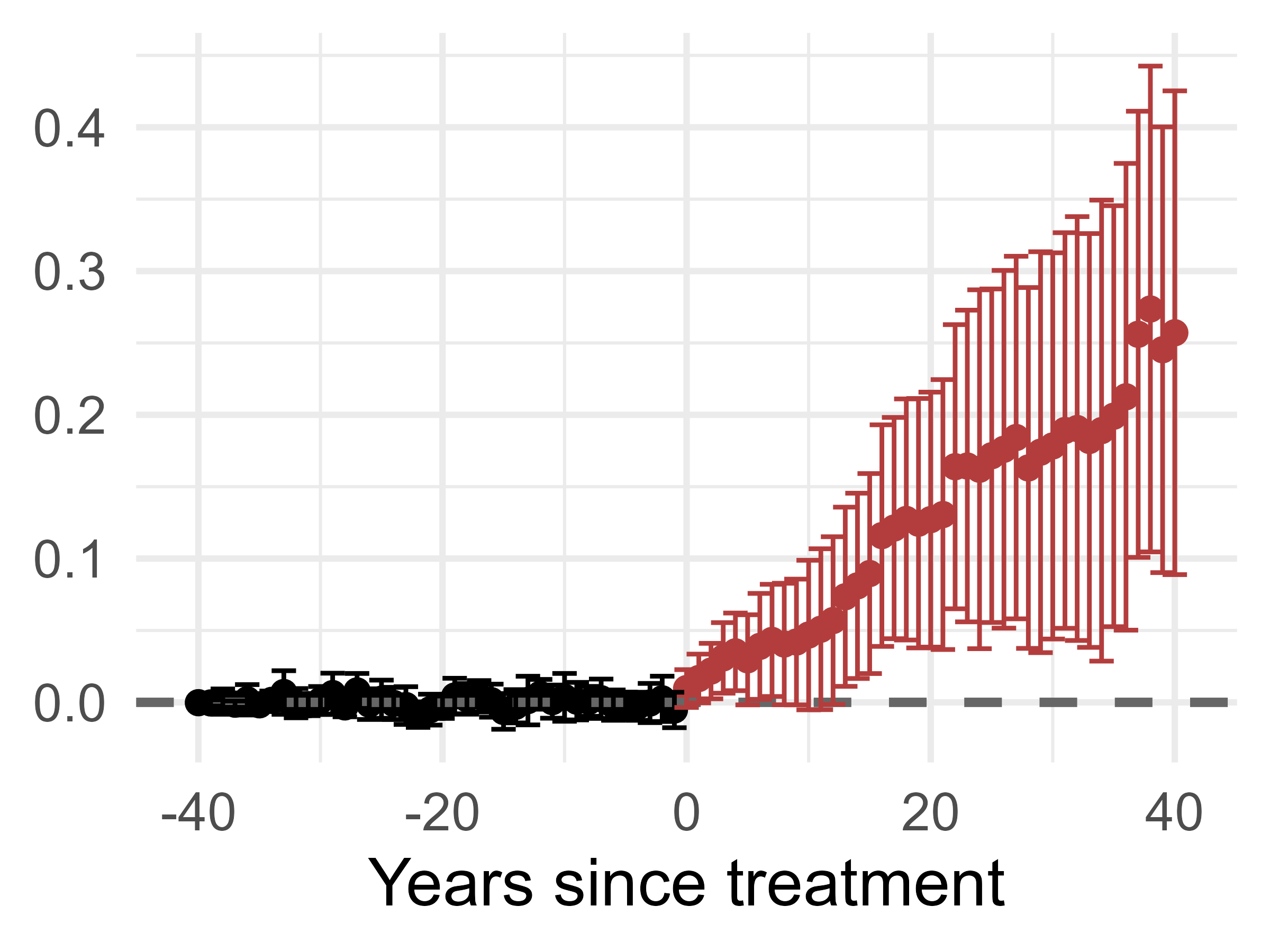}
        \caption*{(C) Community house (density)}
    \end{minipage}
    \hspace{0.5cm}
    \begin{minipage}[b]{0.45\textwidth}
        \centering
        \includegraphics[width=\textwidth]{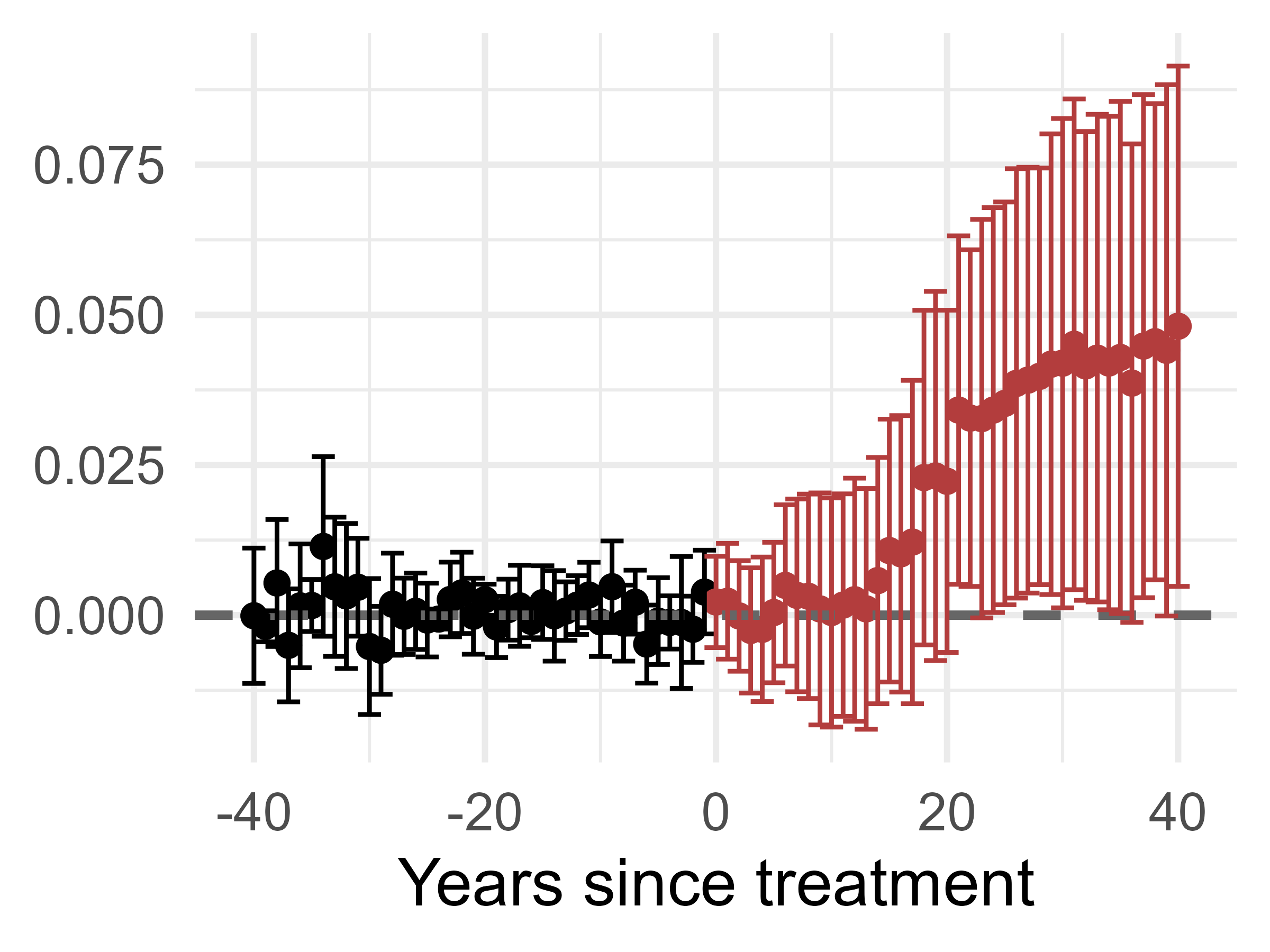}
        \caption*{(D) Folk high school (density)}
    \end{minipage}
    
    \parbox{1\textwidth}{%
        \caption*{\small{\textit{Notes}: Event-study estimates by years since railroad connection for four outcomes capturing Grundtvigian institutions, using the \citet{Callaway2021} estimator relative to never-treated parishes and conditioning on the covariates described in the empirical strategy section. The event study is restricted to event times between -40 and +40 years relative to treatment. Whiskers indicate 95\% confidence intervals.}}
        }
    \label{fig:decompose_grundtvig_dynamic}
\end{figure}

To assess dynamics and the plausibility of parallel trends, Figure~\ref{fig:decompose_grundtvig_dynamic} reports event-study estimates for the Grundtvigian outcomes, analogous to the census analysis. For readability, we restrict the window to event times between -40 and +40 years relative to treatment. The pre-connection coefficients fluctuate closely around zero for both community houses and folk high schools, with no systematic pre-trends. In contrast, significant effects emerge only after railroad connection. This pattern supports the conditional parallel trends assumption for the institutional outcomes as well.

Panel (A) of Figure~\ref{fig:decompose_grundtvig_dynamic} reports dynamic effects on the probability that a parish hosts a community house. The estimates are statistically indistinguishable from zero throughout the window: the confidence intervals include zero at every horizon, both before and after connection. Point estimates rise modestly over the first two decades of exposure and then decline, turning negative at the longest horizons. These movements are imprecisely estimated and never statistically significant, and the confidence intervals widen considerably as exposure lengthens.

Panel (C) instead reports effects on the local density of community houses. In contrast to the establishment dummy in Panel (A), density is flat and close to zero before connection and then rises steadily with years of railroad exposure, becoming statistically distinguishable from zero after roughly fifteen years and continuing to grow thereafter. This pattern is consistent with a diffusion mechanism: rail connection facilitates regional exposure to Grundtvigian ideas, while community houses are established where they are locally useful, often outside the immediate railroad nodes.

For folk high schools, both the probability of establishment (Panel B) and local density (Panel D) increase with additional years of exposure. The establishment effects build up gradually: point estimates remain close to zero during the first decade and then rise steadily, reaching about 4 percentage points after four decades, although the year-by-year estimates are individually insignificant. When aggregated across years, as in the regression table, a small positive post-connection effect is nevertheless detectable. The density effects in Panel D follow a similar gradual build-up, and the confidence intervals begin to exclude zero after roughly twenty years of exposure. In contrast to community houses, far fewer folk high schools were established, and they relied on attracting students from a wider catchment area. Rail connectivity therefore both facilitated the spread of the model and made locations on or near the railroad network more attractive for founding such institutions. Taken together, the event-study patterns reinforce the interpretation that railroads acted as conduits for the diffusion of Grundtvigian institutions, with effects that are more clearly visible in local density than in parish-level siting, especially for community houses. 

The results in this section indicate that railroad connection shaped both economic change and the diffusion of Grundtvigian institutions in Denmark. Rail access increased population, internal migration, and the share of workers in manufacturing and other non-agricultural occupations. At the same time, railroad connection increased local exposure to community houses and folk high schools, and it raised the probability that connected parishes hosted a folk high school. Taken together, the evidence suggests that railroad expansion contributed to Denmark's modernization by reshaping patterns of economic activity, mobility, and cultural organization.

\FloatBarrier
\section{Conclusion}
We have examined the role of railroad expansion in shaping Denmark's economic transformation and the diffusion of Grundtvigian institutions during the nineteenth century. While previous research has documented the economic impact of railroads in facilitating trade, urbanization, and structural change, our study goes further by exploring their role in fostering institutional change. Specifically, we provide evidence that railroads not only contributed to population growth and occupational change but also, though less robustly so, facilitated the diffusion of Grundtvigian institutions linked to civic engagement and the cooperative movement.

Our empirical findings lend support to the idea that railroad access significantly increased local population growth, consistent with findings from other historical contexts. This expansion was not merely redistributive; instead, it facilitated a broader transition away from agriculture and towards manufacturing and services, reflecting structural change that was fundamental to Denmark's modernization. The economic effects appear to have been complemented by institutional changes. Railroad-connected areas showed higher densities of community houses and folk high schools, and a significantly higher probability of hosting a folk high school. These institutional patterns are consistent with railroads acting as conduits for the diffusion of Grundtvigian ideas, with effects that are more visible in regional density than in parish-level siting. These institutions, which were central to the Grundtvigian movement, provided the organizational framework for civic engagement and collective decision-making, key preconditions for the emergence of Denmark's cooperative creameries. 

By providing evidence consistent with a causal link between railroad expansion and the diffusion of Grundtvigian institutions, our findings contribute to a broader literature on the interplay between infrastructure development and institutional change. The tentative sequencing implied by the event-study results---immediate economic effects, more gradual institutional ones---is itself an interesting pattern worth further investigation. The case of Denmark illustrates how improved connectivity can serve as both an economic and cultural catalyst, reinforcing inclusive institutions that promote long-term development.

Moreover, our findings underscore the importance of sequencing in the development process. Railroads preceded the establishment of cooperatives, suggesting that economic modernization created the conditions necessary for institutional innovation. This sequence aligns with broader theories of development, which emphasize the interaction between physical infrastructure, market access, and institutional capacity. In Denmark's case, railroads not only enabled economic growth but also created the social preconditions for cooperative organization, which became a cornerstone of the country's economic model.

These insights have broader implications for understanding how infrastructure investments shape national trajectories. While the Danish case is historically specific, it resonates with contemporary debates on the role of transport networks in fostering inclusive development. Our findings suggest that policymakers should consider not only the economic benefits of infrastructure investments but also their potential to influence institutional change. By fostering civic participation, education, and collective decision-making, infrastructure projects can have far-reaching effects beyond their immediate economic returns.

Taken together, our results highlight that the impact of railroads extended beyond commerce and industry. They helped shape the very institutions that underpinned Denmark's modern economic and social model. Future research could build on this work by further investigating the mechanisms through which transport networks influence institutional evolution, as well as exploring how similar dynamics have played out in other historical and contemporary contexts.

\newpage
\bibliographystyle{apacite}
\bibliography{main}

\section*{Data availability statement}
All code required to reproduce the analysis from raw data through the final regressions is available at \\ \url{https://github.com/christianvedels/Tracks_to_modernity}. The full replication package is also available as "Tracks to Modernity: Railroads, Growth, and Social Movements in Denmark." Inter-university Consortium for Political and Social Research (ICPSR). \url{https://doi.org/10.3886/ICPSR312350.V1}.

\newpage
\setcounter{table}{0}
\setcounter{figure}{0}
\setcounter{section}{0}
\renewcommand*{\thesection}{\Alph{section}}
\renewcommand{\thefigure}{A\arabic{figure}}
\renewcommand{\thetable}{A\arabic{table}}
\pagenumbering{roman}

\begin{appendices}

\begin{titlepage}

    \begin{center}
        \Large
        \textbf{Appendix:} \\
        Tracks to Modernity: Railroads, Growth, and Civic Institutions in Denmark \\
        \vspace{0.5cm}
        \large
        
        Tom Görges, TU Dortmund University \\ 
        Magnus Ørberg Rove, Statistics Denmark\\
        Paul Sharp, University of Southern Denmark \\ 
        Christian Vedel, University of Southern Denmark \\

        \vspace{1cm}

        \normalsize
        \begin{minipage}{0.80\textwidth}
    
            \setcounter{tocdepth}{1}
            \tableofappendices

        \end{minipage}
    
        \vfill

    \end{center}

\end{titlepage}

\section{Density plots by year} \label{densities_by_year}
\FloatBarrier

\begin{figure}[ht]
    \centering
    \caption{Distribution of variables in 1850 depending on period in which the parish was connected to the railroad}
    \includegraphics[width=1\linewidth]{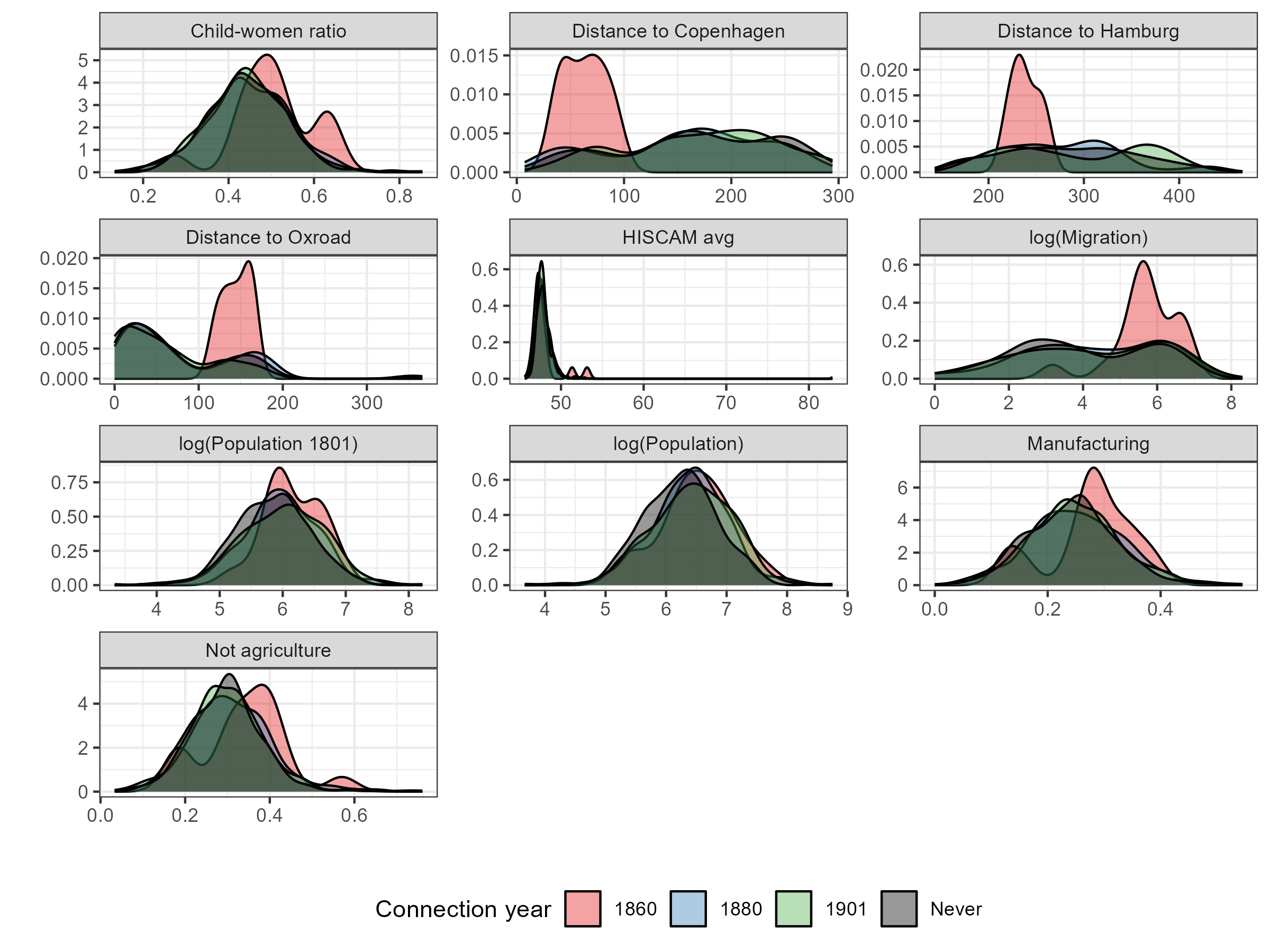}
    \parbox{1\textwidth}{\caption*{\small{\textit{Notes}: Density estimates of a number of variables from the census data and other covariates, separated by the year when each parish was connected to the railroad. Parishes that were already connected in 1850 (8 out of 1589) are excluded from these plots. }}}\label{fig:densities_by_treat_year}
\end{figure}

The density plots shown in Figure \ref{fig:densities_by_treat_year} offer a visual comparison of key variables from the census data, grouped by the period during which each parish was connected to the railroad. These groups consist of parishes connected to the railroad before 1860, between 1861 and 1880, between 1881 and 1901, and those that were never connected. The distinction between the periods of connection helps us assess whether there were systematic differences between parishes that received railroad connections at different points in time. 

The pre-1860 group is exceptional: it consists only of a handful parishes along the second Danish main line from Roskilde to Korsør. Consequently, the corresponding distance measures differ sharply from the later cohorts in a largely mechanical way. By contrast, the distributions of the remaining variables overlap substantially across cohorts, indicating that the pre-1860 group does not differ systematically from later-connected parishes outside of the distance measures. Migration is a partial exception, where the earliest-connected parishes exhibit more pronounced differences. Overall, the evidence points to meaningful overlap in the main variables across cohorts, with the largest cohort differences concentrated in distance measures for historical and geographic reasons.

\clearpage
\section{Balance and distribution tests in 1850}\label{appendix:balance}
\FloatBarrier

Table~\ref{tbl:balance_tests_ever_treated} reports pre-treatment balance and distribution tests using the 1850 cross section. We restrict attention to parishes that are not yet connected in 1850 and classify them by whether they are ever connected later. The "Ever Connected" column reports the coefficient from an OLS regression of each variable on an indicator for eventual connection; this is equivalent to a difference-in-means (t-test) between ever- and never-connected parishes. The final column reports Kolmogorov--Smirnov (KS) tests of equality of the full distributions.

\begin{table}[ht]
\caption{Regression Tests: Ever Connected vs Never Connected}
\centering
\begin{tabular}{lcccc}
\toprule
Variable & Mean & SD & Ever Connected & KS D-stat\\
\midrule
log(Population) & 6.315 & 0.626 & 0.167*** & 0.137***\\
 &  &  & (0.032) & \\
Child-women ratio & 0.444 & 0.094 & 0.007 & 0.039\\
 &  &  & (0.005) & \\
Manufacturing & 0.245 & 0.080 & 0.010** & 0.076**\\
 &  &  & (0.004) & \\
Not agriculture & 0.299 & 0.089 & 0.012*** & 0.083***\\
 &  &  & (0.005) & \\
HISCAM avg & 47.872 & 1.264 & 0.004 & 0.055\\
 &  &  & (0.064) & \\
log(Migration) & 4.112 & 1.876 & 0.315*** & 0.102***\\
 &  &  & (0.097) & \\
Distance to Hamburg & 283.734 & 69.693 & -1.543 & 0.056\\
 &  &  & (3.541) & \\
Distance to Copenhagen & 163.897 & 73.447 & -6.501* & 0.065*\\
 &  &  & (3.728) & \\
Distance to Oxroad & 71.853 & 64.757 & 1.442 & 0.030\\
 &  &  & (3.291) & \\
log(Population 1801) & 5.912 & 0.603 & 0.133*** & 0.113***\\
 &  &  & (0.030) & \\
\bottomrule
\end{tabular}
\parbox{0.99\textwidth}{\caption*{\small{\textit{Notes}: Mean and standard deviation of each variable in 1850. ``Ever Connected'' reports coefficients from OLS regressions of each variable on an indicator for eventual railroad connection (1=ever connected, 0=never connected); standard errors in parentheses. ``KS D-stat'' reports the Kolmogorov--Smirnov test statistic for equality of distributions. Significance: * $p<0.10$, ** $p<0.05$, *** $p<0.01$.}}}
\label{tbl:balance_tests_ever_treated}
\end{table}

\FloatBarrier

\section{Excluding Early-Connected Parishes}\label{appendix:excl_early}

The balance tests in Appendix~\ref{appendix:balance} and the density plots in Appendix~\ref{densities_by_year} indicate that differences between connected and never-connected parishes are largely driven by the small set of parishes connected very early. As a robustness check, Tables~\ref{tbl:rail_census_excl_early} and~\ref{tbl:rail_grundtvig_excl_early} therefore re-estimate the baseline specifications with controls after excluding all parishes connected by 1860 (25 parishes), so that treatment effects are identified only from comparisons between later-connected and never-connected parishes. The results are similar to the baseline estimates.

\begin{landscape}
    \begin{table}[ht]
    \caption{Railroads and Local Development (excluding parishes connected by 1860)}
    \centering
    \begin{tabular}{lcccccc}
  \toprule
  Outcome: & log(Pop.) & Child-women ratio & Manufacturing & Not Agriculture & HISCAM avg & log(Migration) \\
           & (1) & (2) & (3) & (4) & (5) & (6) \\
  \midrule
  \multicolumn{7}{l}{\textbf{A. TWFE estimates}}\\
  Connected railroad &  0.0675$^{***}$ & 0.0010$^{}$ & 0.0177$^{***}$ & 0.0199$^{***}$ & 0.1022$^{}$ & 0.1337$^{***}$  \\
                    &  (0.0096) & (0.0048) & (0.0032) & (0.0040) & (0.0630) & (0.0404)  \\
  \cmidrule(lr){2-7}
  Observations      &  6256 & 6236 & 6246 & 6246 & 6255 & 6164  \\
  Mean of outcome   &  6.4708 & 0.4882 & 0.2552 & 0.3490 & 48.6178 & 4.3963  \\
  \midrule
  \multicolumn{7}{l}{\textbf{B. Callaway and Sant'Anna estimates}}\\
  Connected railroad &  0.0716$^{***}$ & 0.0047$^{}$ & 0.0181$^{***}$ & 0.0226$^{***}$ & 0.1393$^{*}$ & 0.1111$^{**}$  \\
                    &  (0.0106) & (0.0050) & (0.0038) & (0.0044) & (0.0772) & (0.0467)  \\
  \cmidrule(lr){2-7}
  Observations      &  6252 & 6172 & 6212 & 6212 & 6248 & 5932  \\
  Mean of outcome   &  6.4703 & 0.4882 & 0.2551 & 0.3486 & 48.6169 & 4.4728  \\
  \bottomrule
\end{tabular}

    \parbox{1.0\textwidth}{\caption*{\small{\textit{Notes}: This table replicates the baseline specifications with controls, excluding all parishes connected to the railroad by 1860. Standard errors clustered at the parish level in parentheses. *** $p<0.01$ ** $p<0.05$ * $p<0.10$}}}\label{tbl:rail_census_excl_early}
    \end{table}
\end{landscape}

\begin{table}[ht]
\caption{Railroads and Grundtvigianism (excluding parishes connected by 1860)}
\centering
\footnotesize
\resizebox{\textwidth}{!}{%
\begin{tabular}{lcccc}
  \toprule
  Outcome: & Community house & Folk high school & \makecell{Density Community \\ houses (MA)} & \makecell{Density Folk High \\ Schools (MA)} \\
           & (1) & (2) & (3) & (4)  \\
  \midrule
  \multicolumn{5}{l}{\textbf{A. TWFE estimates}}\\
  Connected railroad &  0.0345$^{**}$ & 0.0114$^{*}$ & 0.0751$^{***}$ & 0.0134$^{**}$  \\
                    &  (0.0138) & (0.0063) & (0.0273) & (0.0063)  \\
  \cmidrule(lr){2-5}
  Observations      &  117300 & 117300 & 117300 & 117300  \\
  Mean of outcome   &  0.1732 & 0.0297 & 4.4716 & 0.7021  \\
  \midrule
  \multicolumn{5}{l}{\textbf{B. Callaway and Sant'Anna estimates}}\\
  Connected railroad &  0.0022$^{}$ & 0.0210$^{***}$ & 0.1365$^{***}$ & 0.0247$^{***}$  \\
                    &  (0.0153) & (0.0077) & (0.0249) & (0.0082)  \\
  \cmidrule(lr){2-5}
  Observations      &  117225 & 117225 & 117225 & 117225  \\
  Mean of outcome   &  0.1734 & 0.0297 & 4.4722 & 0.7022  \\
  \bottomrule
\end{tabular}
}

\parbox{0.99\textwidth}{\caption*{\small{\textit{Notes}: This table replicates the baseline specifications with controls, excluding all parishes connected to the railroad by 1860. Standard errors clustered at the parish level in parentheses. *** $p<0.01$ ** $p<0.05$ * $p<0.10$}}}\label{tbl:rail_grundtvig_excl_early}
\end{table}

\FloatBarrier

\section{Alternative Standard Errors}\label{Alt_ses_census_and_grundtvig}

To address spatial dependence in treatment assignment and outcomes, we report (i) standard errors clustered at the county level and (ii) Conley (spatial HAC) standard errors for the TWFE specifications. For Conley, we report results using 10 km, 25 km, and 50 km cutoff radii.

\subsection*{County-level clustering}

\begin{landscape}
    \begin{table}[ht]
    \caption{Railroads and Local Development (SEs clustered at the county level)}
    \centering
    \begin{tabular}{lcccccc}
  \toprule
  Outcome: & log(Pop.) & Child-women ratio & Manufacturing & Not Agriculture & HISCAM avg & log(Migration) \\
           & (1) & (2) & (3) & (4) & (5) & (6) \\
  \midrule
  \multicolumn{7}{l}{\textbf{A. TWFE estimates}}\\
  Connected railroad &  0.0664$^{***}$ & 0.0016$^{}$ & 0.0173$^{***}$ & 0.0186$^{***}$ & 0.0862$^{}$ & 0.1353$^{***}$  \\
                    &  (0.0089) & (0.0050) & (0.0030) & (0.0039) & (0.0657) & (0.0459)  \\
  \cmidrule(lr){2-7}
  Observations      &  6356 & 6336 & 6346 & 6346 & 6355 & 6264  \\
  Mean of outcome   &  6.4738 & 0.4883 & 0.2559 & 0.3496 & 48.6308 & 4.4198  \\
  \midrule
  \multicolumn{7}{l}{\textbf{B. Callaway and Sant'Anna estimates}}\\
  Connected railroad &  0.0703$^{***}$ & 0.0050$^{}$ & 0.0177$^{***}$ & 0.0203$^{***}$ & 0.1252$^{}$ & 0.0998$^{**}$  \\
                    &  (0.0108) & (0.0050) & (0.0037) & (0.0048) & (0.0942) & (0.0454)  \\
  \cmidrule(lr){2-7}
  Observations      &  6320 & 6240 & 6280 & 6280 & 6316 & 6000  \\
  Mean of outcome   &  6.4716 & 0.4884 & 0.2555 & 0.3489 & 48.6185 & 4.4884  \\
  \bottomrule
\end{tabular}

    \parbox{1.0\textwidth}{\caption*{\small{\textit{Notes}: Standard errors clustered at the county level in parentheses. *** $p<0.01$ ** $p<0.05$ * $p<0.10$}}}\label{tbl:rail_census_alt_ses_county}
    \end{table}  
\end{landscape}

\begin{landscape}
    \begin{table}[ht]
    \caption{Railroads and Grundtvigianism (SEs clustered at the county level)}
    \centering
    \resizebox{\textwidth}{!}{%
\begin{tabular}{lcccc}
  \toprule
  Outcome: & Community house & Folk high school & \makecell{Density Community \\ houses (MA)} & \makecell{Density Folk High \\ Schools (MA)} \\
           & (1) & (2) & (3) & (4)  \\
  \midrule
  \multicolumn{5}{l}{\textbf{A. TWFE estimates}}\\
  Connected railroad &  0.0357$^{*}$ & 0.0108$^{*}$ & 0.0768$^{*}$ & 0.0125$^{}$  \\
                    &  (0.0179) & (0.0060) & (0.0381) & (0.0089)  \\
  \cmidrule(lr){2-5}
  Observations      &  119175 & 119175 & 119175 & 119175  \\
  Mean of outcome   &  0.1715 & 0.0295 & 4.4639 & 0.7024  \\
  \midrule
  \multicolumn{5}{l}{\textbf{B. Callaway and Sant'Anna estimates}}\\
  Connected railroad &  0.0037$^{}$ & 0.0168$^{}$ & 0.1450$^{***}$ & 0.0188$^{*}$  \\
                    &  (0.0176) & (0.0103) & (0.0395) & (0.0107)  \\
  \cmidrule(lr){2-5}
  Observations      &  119100 & 119100 & 119100 & 119100  \\
  Mean of outcome   &  0.1716 & 0.0295 & 4.4646 & 0.7026  \\
  \bottomrule
\end{tabular}
}

    \parbox{1.0\textwidth}{\caption*{\small{\textit{Notes}: Standard errors clustered at the county level in parentheses. *** $p<0.01$ ** $p<0.05$ * $p<0.10$}}}\label{tbl:rail_grundtvig_alt_ses_county}
    \end{table}  
\end{landscape}

\subsection*{Conley (spatial HAC) standard errors}

\begin{landscape}
    \begin{table}[ht]
    \caption{Railroads and Local Development (Conley SEs, 10 km cutoff)}
    \centering
    \begin{tabular}{lcccccc}
  \toprule
  Outcome: & log(Pop.) & Child-women ratio & Manufacturing & Not Agriculture & HISCAM avg & log(Migration) \\
           & (1) & (2) & (3) & (4) & (5) & (6) \\
  \midrule
  Connected railroad &  0.0664$^{***}$ & 0.0016$^{}$ & 0.0173$^{***}$ & 0.0186$^{***}$ & 0.0862$^{}$ & 0.1353$^{***}$  \\
                    &  (0.0104) & (0.0049) & (0.0036) & (0.0039) & (0.0600) & (0.0471)  \\
  \cmidrule(lr){2-7}
  Observations      &  6356 & 6336 & 6346 & 6346 & 6355 & 6264  \\
  Mean of outcome   &  6.4738 & 0.4883 & 0.2559 & 0.3496 & 48.6308 & 4.4198  \\
  \bottomrule
\end{tabular}

    \parbox{1.0\textwidth}{\caption*{\small{\textit{Notes}: TWFE estimates with Conley (spatial HAC) standard errors using a 10 km cutoff radius. *** $p<0.01$ ** $p<0.05$ * $p<0.10$}}}\label{tbl:rail_dev_control_conley10}
    \end{table}  
\end{landscape}

\begin{table}[ht]
\caption{Railroads and Grundtvigianism (Conley SEs, 10 km cutoff)}
\centering
\resizebox{\textwidth}{!}{%
\begin{tabular}{lcccc}
  \toprule
  Outcome: & Community house & Folk high school & \makecell{Density Community \\ houses (MA)} & \makecell{Density Folk High \\ Schools (MA)} \\
           & (1) & (2) & (3) & (4)  \\
  \midrule
  Connected railroad &  0.0357$^{**}$ & 0.0108$^{}$ & 0.0768$^{**}$ & 0.0125$^{}$  \\
                    &  (0.0159) & (0.0066) & (0.0334) & (0.0077)  \\
  \cmidrule(lr){2-5}
  Observations      &  119175 & 119175 & 119175 & 119175  \\
  Mean of outcome   &  0.1715 & 0.0295 & 4.4639 & 0.7024  \\
  \bottomrule
\end{tabular}
}

\parbox{0.99\textwidth}{\caption*{\small{\textit{Notes}: TWFE estimates with Conley (spatial HAC) standard errors using a 10 km cutoff radius. *** $p<0.01$ ** $p<0.05$ * $p<0.10$}}}\label{tbl:rail_grundt_control_conley10}
\end{table}

\begin{landscape}
    \begin{table}[ht]
    \caption{Railroads and Local Development (Conley SEs, 25 km cutoff)}
    \centering
    \begin{tabular}{lcccccc}
  \toprule
  Outcome: & log(Pop.) & Child-women ratio & Manufacturing & Not Agriculture & HISCAM avg & log(Migration) \\
           & (1) & (2) & (3) & (4) & (5) & (6) \\
  \midrule
  Connected railroad &  0.0664$^{***}$ & 0.0016$^{}$ & 0.0173$^{***}$ & 0.0186$^{***}$ & 0.0862$^{}$ & 0.1353$^{**}$  \\
                    &  (0.0119) & (0.0069) & (0.0041) & (0.0052) & (0.0563) & (0.0542)  \\
  \cmidrule(lr){2-7}
  Observations      &  6356 & 6336 & 6346 & 6346 & 6355 & 6264  \\
  Mean of outcome   &  6.4738 & 0.4883 & 0.2559 & 0.3496 & 48.6308 & 4.4198  \\
  \bottomrule
\end{tabular}

    \parbox{1.0\textwidth}{\caption*{\small{\textit{Notes}: TWFE estimates with Conley (spatial HAC) standard errors using a 25 km cutoff radius. *** $p<0.01$ ** $p<0.05$ * $p<0.10$}}}\label{tbl:rail_dev_control_conley25}
    \end{table}  
\end{landscape}

\begin{table}[ht]
\caption{Railroads and Grundtvigianism (Conley SEs, 25 km cutoff)}
\centering
\resizebox{\textwidth}{!}{%
\begin{tabular}{lcccc}
  \toprule
  Outcome: & Community house & Folk high school & \makecell{Density Community \\ houses (MA)} & \makecell{Density Folk High \\ Schools (MA)} \\
           & (1) & (2) & (3) & (4)  \\
  \midrule
  Connected railroad &  0.0357$^{*}$ & 0.0108$^{}$ & 0.0768$^{*}$ & 0.0125$^{}$  \\
                    &  (0.0199) & (0.0085) & (0.0394) & (0.0113)  \\
  \cmidrule(lr){2-5}
  Observations      &  119175 & 119175 & 119175 & 119175  \\
  Mean of outcome   &  0.1715 & 0.0295 & 4.4639 & 0.7024  \\
  \bottomrule
\end{tabular}
}

\parbox{0.99\textwidth}{\caption*{\small{\textit{Notes}: TWFE estimates with Conley (spatial HAC) standard errors using a 25 km cutoff radius. *** $p<0.01$ ** $p<0.05$ * $p<0.10$}}}\label{tbl:rail_grundt_control_conley25}
\end{table}

\begin{landscape}
    \begin{table}[ht]
    \caption{Railroads and Local Development (Conley SEs, 50 km cutoff)}
    \centering
    \begin{tabular}{lcccccc}
  \toprule
  Outcome: & log(Pop.) & Child-women ratio & Manufacturing & Not Agriculture & HISCAM avg & log(Migration) \\
           & (1) & (2) & (3) & (4) & (5) & (6) \\
  \midrule
  Connected railroad &  0.0664$^{***}$ & 0.0016$^{}$ & 0.0173$^{***}$ & 0.0186$^{***}$ & 0.0862$^{}$ & 0.1353$^{**}$  \\
                    &  (0.0159) & (0.0085) & (0.0052) & (0.0062) & (0.0875) & (0.0690)  \\
  \cmidrule(lr){2-7}
  Observations      &  6356 & 6336 & 6346 & 6346 & 6355 & 6264  \\
  Mean of outcome   &  6.4738 & 0.4883 & 0.2559 & 0.3496 & 48.6308 & 4.4198  \\
  \bottomrule
\end{tabular}

    \parbox{1.0\textwidth}{\caption*{\small{\textit{Notes}: TWFE estimates with Conley (spatial HAC) standard errors using a 50 km cutoff radius. *** $p<0.01$ ** $p<0.05$ * $p<0.10$}}}\label{tbl:rail_dev_control_conley}
    \end{table}  
\end{landscape}

\begin{landscape}
    \begin{table}[ht]
    \caption{Railroads and Grundtvigianism (Conley SEs, 50 km cutoff)}
    \centering
    \resizebox{\textwidth}{!}{%
\begin{tabular}{lcccc}
  \toprule
  Outcome: & Community house & Folk high school & \makecell{Density Community \\ houses (MA)} & \makecell{Density Folk High \\ Schools (MA)} \\
           & (1) & (2) & (3) & (4)  \\
  \midrule
  Connected railroad &  0.0357$^{}$ & 0.0108$^{}$ & 0.0768$^{}$ & 0.0125$^{}$  \\
                    &  (0.0222) & (0.0094) & (0.0564) & (0.0126)  \\
  \cmidrule(lr){2-5}
  Observations      &  119175 & 119175 & 119175 & 119175  \\
  Mean of outcome   &  0.1715 & 0.0295 & 4.4639 & 0.7024  \\
  \bottomrule
\end{tabular}
}

    \parbox{1.0\textwidth}{\caption*{\small{\textit{Notes}: TWFE estimates with Conley (spatial HAC) standard errors using a 50 km cutoff radius. *** $p<0.01$ ** $p<0.05$ * $p<0.10$}}}\label{tbl:rail_grundt_control_conley}
    \end{table}  
\end{landscape}

\FloatBarrier

\section{Specifications without Covariates} \label{regresssionsnocontrols}

Tables \ref{tbl:rail_census_nocontrol} and \ref{tbl:rail_grundtvig_nocontrol} present estimates of the effect of railroad access on local development and Grundtvigian institutions, now without controls. 

\begin{landscape}
    \begin{table}[ht]
    \caption{Railroads and Local Development (without controls)}
    \centering
    \begin{tabular}{lcccccc}
  \toprule
  Outcome: & log(Pop.) & Child-women ratio & Manufacturing & Not Agriculture & HISCAM avg & log(Migration) \\
           & (1) & (2) & (3) & (4) & (5) & (6) \\
  \midrule
  \multicolumn{7}{l}{\textbf{A. TWFE estimates}}\\
  Connected railroad &  0.0654$^{***}$ & -0.0005$^{}$ & 0.0173$^{***}$ & 0.0180$^{***}$ & 0.1723$^{***}$ & 0.1482$^{***}$  \\
                    &  (0.0106) & (0.0048) & (0.0031) & (0.0038) & (0.0605) & (0.0411)  \\
  \cmidrule(lr){2-7}
  Observations      &  6356 & 6336 & 6346 & 6346 & 6355 & 6264  \\
  Mean of outcome   &  6.4738 & 0.4883 & 0.2559 & 0.3496 & 48.6308 & 4.4198  \\
  \midrule
  \multicolumn{7}{l}{\textbf{B. Callaway and Sant'Anna estimates}}\\
  Connected railroad &  0.0656$^{***}$ & 0.0010$^{}$ & 0.0176$^{***}$ & 0.0185$^{***}$ & 0.2122$^{***}$ & 0.1086$^{**}$  \\
                    &  (0.0111) & (0.0049) & (0.0034) & (0.0042) & (0.0680) & (0.0436)  \\
  \cmidrule(lr){2-7}
  Observations      &  6324 & 6244 & 6284 & 6284 & 6320 & 6004  \\
  Mean of outcome   &  6.4722 & 0.4884 & 0.2556 & 0.3491 & 48.6194 & 4.4896  \\
  \bottomrule
\end{tabular}

    \parbox{1.0\textwidth}{\caption*{\small{\textit{Notes}: Clustered (Parish) standard errors in parenthesis *** $p< 0.01$ ** $p< 0.05$ * $p< 0.10$}}}\label{tbl:rail_census_nocontrol}
    \end{table}  
\end{landscape}

\begin{table}[ht]
\caption{Railroads and Grundtvigianism (without controls)}
\centering
\footnotesize
\resizebox{\textwidth}{!}{%
\begin{tabular}{lcccc}
  \toprule
  Outcome: & Community house & Folk high school & \makecell{Density Community \\ houses (MA)} & \makecell{Density Folk High \\ Schools (MA)} \\
           & (1) & (2) & (3) & (4)  \\
  \midrule
  \multicolumn{5}{l}{\textbf{A. TWFE estimates}}\\
  Connected railroad &  0.0409$^{***}$ & 0.0106$^{*}$ & -0.0254$^{}$ & 0.0143$^{**}$  \\
                    &  (0.0134) & (0.0058) & (0.0696) & (0.0072)  \\
  \cmidrule(lr){2-5}
  Observations      &  119175 & 119175 & 119175 & 119175  \\
  Mean of outcome   &  0.1715 & 0.0295 & 4.4639 & 0.7024  \\
  \midrule
  \multicolumn{5}{l}{\textbf{B. Callaway and Sant'Anna estimates}}\\
  Connected railroad &  0.0056$^{}$ & 0.0171$^{**}$ & 0.0578$^{}$ & 0.0315$^{***}$  \\
                    &  (0.0131) & (0.0068) & (0.0727) & (0.0074)  \\
  \cmidrule(lr){2-5}
  Observations      &  119175 & 119175 & 119175 & 119175  \\
  Mean of outcome   &  0.1715 & 0.0295 & 4.4639 & 0.7024  \\
  \bottomrule
\end{tabular}
}

\parbox{0.99\textwidth}{\caption*{\small{\textit{Notes}: Measures of Grundtvigianism and the spread of the railroad. Clustered (Parish) standard errors in parenthesis *** $p< 0.01$ ** $p< 0.05$ * $p< 0.10$}}}\label{tbl:rail_grundtvig_nocontrol}
\end{table}

\end{appendices}

\end{document}